\documentclass[10pt,aps,prd,amsmath,amssymb,nofootinbib,superscriptaddress,preprintnumbers,hidelinks,
nobibnotes,
floatfix
]{revtex4-2}

\usepackage{orcidlink}
\usepackage{graphicx}% Include figure files
\usepackage{dcolumn}% Align table columns on decimal point
\usepackage{bm}% bold math
\usepackage{cancel}
\usepackage{physics}
\usepackage{caption,subcaption}
\usepackage{booktabs}
\usepackage{multirow}
\usepackage{fancyhdr}
\usepackage{tikz}
\usetikzlibrary{calc}

\newcommand{\xh}{x_H}

\newcommand{\rps}{r_\text{ps}}

\newcommand{\bsh}{b_\text{sh}}

\newcommand{\hr}{r_H}

\newcommand{\rmin}{r_\text{min}}
\newcommand{\robs}{r_\text{O}}

\newcommand{\zh}{r_H}

\newcommand{\fn}[1]{f^{(#1)}_{H}}
\newcommand{\hn}[1]{h^{(#1)}_{H}}

\newcommand{\co}{r}
\newcommand{\ma}{M}

\newcommand{\rd}{\kappa}

\newcommand{\uva}[1]{u_{#1}}

\begin{document}

%\preprint{.......}

%\title{Stepping away from the abyss}

\title{\Large \bf Black Hole Shadow and other Observables away from the Horizon:\\[5pt]
Extending the Effective Metric Descriptions
}
\author{Manuel Del Piano {\Large \orcidlink{0000-0003-4515-8787}}\,}
\email[]{manuel.delpiano-ssm@unina.it}
%\email[]{\href{mailto:manuel.delpiano-ssm@unina.it}{manuel.delpiano-ssm@unina.it}}
\affiliation{Scuola Superiore Meridionale, Largo S. Marcellino, 10, 80138 Napoli, Italy}
\affiliation{INFN sezione di Napoli, via Cintia, 80126 Napoli, Italy}
\affiliation{Quantum  Theory Center ($\hbar$QTC) \& D-IAS, Southern Denmark Univ., Campusvej 55, 5230 Odense M, Denmark}

\author{Stefan Hohenegger {\Large \orcidlink{0000-0001-6564-0795}}\,}
\email[]{s.hohenegger@ipnl.in2p3.fr}
\affiliation{Univ Lyon, Univ Claude Bernard Lyon 1, CNRS/IN2P3, IP2I Lyon, UMR 5822, F-69622, Villeurbanne, France}

\author{Francesco Sannino {\Large\orcidlink{0000-0003-2361-5326}}\,}
\email[]{sannino@qtc.sdu.dk}
\affiliation{Scuola Superiore Meridionale, Largo S. Marcellino, 10, 80138 Napoli, Italy}
\affiliation{INFN sezione di Napoli, via Cintia, 80126 Napoli, Italy}
\affiliation{Quantum Theory Center ($\hbar$QTC) \& D-IAS, Southern Denmark Univ., Campusvej 55, 5230 Odense M, Denmark}
\affiliation{Dept. of Physics E. Pancini, Università di Napoli Federico II, via Cintia, 80126 Napoli, Italy}

\begin{abstract}
In previous work \cite{DelPiano:2023fiw,DelPiano:2024gvw} we have developed a model-independent, effective description of quantum deformed, spherically symmetric and static black holes in four dimensions. The deformations of the metric are captured by two functions of the physical distance to the horizon, which are provided in the form of self-consistent Taylor series expansions. While this approach efficiently captures physical observables in the immediate vicinity of the horizon, it is expected to encounter problems of convergence at further distances. Therefore, we demonstrate in this paper how to use Padé approximants to extend the range of applicability of this framework. We provide explicit approximations of physical observables that depend on finitely many effective parameters of the deformed black hole geometry, depending on the order of the Padé approximant. By taking the asymptotic limit of this order, we in particular provide a closed-form expression for the black hole shadow of the (fully) deformed geometry, which captures the leading quantum corrections. We illustrate our results for a number of quantum black holes previously proposed in the literature and find that our effective approach provides excellent approximations in all cases.
 \end{abstract}

\maketitle

\section*{Introduction}
Since the initial discovery of the celebrated Schwarzschild solution in General Relativity (GR) \cite{Schwarzschild:1916uq,Wald:1984rg,Maggiore:2018sht,Chandrasekhar:1985kt,Carroll:2004st,misner1973gravitation}, \emph{black holes} have been a central object of theoretical study for almost a century. In particular in the context of (quantum) modified theories of gravity (that go beyond the classical theory), numerous space-time geometries that resemble black holes have been proposed theoretically and their properties studied in detail. With the advent of large scale astrophysical experiments (such as for example the Event Horizon Telescope \cite{EventHorizonTelescope:2019dse, EventHorizonTelescope:2021dqv, EventHorizonTelescope:2022wkp, EventHorizonTelescope:2022xqj,Vagnozzi:2022moj} or the detection of gravitational waves stemming from black hole binaries and mergers \cite{LIGOScientific:2020tif,LIGOScientific:2020stg,LIGOScientific:2018dkp,LIGOScientific:2016dsl,LIGOScientific:2016aoc}), black holes have also moved to the forefront of observational studies. This has sparked the hope that (with an expected increase in the precision of the experimental measurements), black holes could reveal traces of physics beyond GR that might provide clues towards a theory of Quantum Gravity.\\

As there is currently no universally accepted such theory, the correct description of black holes beyond GR is not at all obvious. In this work, we shall work under the assumption, that they still allow a geometric description in the form of a \emph{metric}, at least in the region outside of the event horizon, which is accessible to direct observations. Since furthermore quantum\footnote{In the following we shall call modifications of black holes in vacuum that go beyond classical General Relativity 'quantum' corrections. While we indeed have mostly in mind effects that are due to a quantum theory of gravity, our approach does not hinge on the quantum nature of these effects and our results are equally applicable to any other modified theory of gravity.} effects are expected to be very small, quantum black holes are usually formulated as deformations of (established) classical solutions: indeed, numerous examples of such deformed geometries have been proposed in the literature, either within the framework of a concrete proposal of a theory of quantum gravity or motivated by symmetries and/or other properties. This plethora of isolated models makes a direct comparison to experimental results as well as a theoretical understanding of universal properties of quantum black holes difficult. Indeed, findings of either type are specific to a concrete (class of) model(s) and difficult to generalise to a larger context: \emph{e.g.} in \cite{Cadoni:2022vsn}, bounds have been formulated on the size of the deformation parameter of a specific black hole, based on astrophysical observations. However, these bounds are a priori only meaningful within the concrete model that was used to describe the deformation of the geometry beyond the Schwarzschild metric.\\

For this reason we have developed in previous works \cite{Binetti:2022xdi,DelPiano:2023fiw,DelPiano:2024gvw} an \emph{effective metric description} (EMD) of black hole geometries that is capable of accommodating the various isolated models, while at the same time capturing universal features of black hole physics. Focusing on spherically symmetric and static black holes, the main feature of these EMD's is to formulate the deformations with respect to the Schwarzschild geometry as functions of a physical quantity. While the choice of the latter does not impact physical observables (in the sense that EMD's based on different choices are equivalent \cite{DelPiano:2024gvw}), previous works \cite{DelPiano:2023fiw} have been based on the \emph{proper distance} $\rho$ to the event horizon of the black hole: we have developed a framework that allows to compute this proper distance in a self-consistent fashion from the metric (which itself depends on $\rho$) in the form of a Taylor series around $\rho=0$. This in turn allows to compute observables, such as the Hawking temperature, that only depend on the geometry at the event horizon of the black hole, as functions of finitely many\footnote{Three in the case of the Hawking temperature.}, universally definable parameters. The latter appear as effective parameters in the Taylor series expansion of the metric deformations in terms of the physical quantity $\rho$. We also mention that using an asymptotic version of the EMDs \cite{Binetti:2022xdi,DAlise:2023hls}, in \cite{Li:2023djs} the precession of bound orbits around the black hole have been studied, while in  \cite{Belfiglio:2024qsa} a version of entanglement entropy was discussed.

The large scale astrophysical experiments mentioned above are in general poised to measure other observables that depend on the black hole geometry at a larger, but not asymptotic, distance to the event horizon. For example, the Event Horizon Telescope is able to measure the \emph{black hole shadow}~\cite{EventHorizonTelescope:2019dse,EventHorizonTelescope:2021dqv,EventHorizonTelescope:2022wkp,EventHorizonTelescope:2022xqj}, which is related to the last stable photon orbit of the black hole \cite{Perlick:2021aok,Carroll:2004st,misner1973gravitation,Bozza:2002zj}. While, relatively speaking, still close to the horizon\footnote{In the classical Schwarzschild geometry, the stable photon orbit is located at $\rps=\frac{3}{2}\,\zh$, where $\zh$ is the position of the event horizon.}, the shadow probes the geometry in a region, where the distance EMD may develop technical problems due to the lack of convergence of the Taylor series expansion inherent in the metric deformations. In this paper, we therefore explore ways to extend the validity of the (distance) EMD to regions that are further away from the event horizon, with the concrete goal to provide (an approximate) expression for the black hole shadow.\\

The core idea is to replace the metric deformations, which in the framework developed in \cite{Binetti:2022xdi,DelPiano:2023fiw,DelPiano:2024gvw} are only given in the form of Taylor series expansions in close proximity of the horizon, by a function that is well defined in a region outside of the black hole. In particular, this region may extend outside of the radius of convergence of the initial Taylor series expansion, thus effectively extending the EMD beyond its original validity, albeit only approximately. These extended EMD's in turn allow us to compute (approximations) of observables that probe a region further away from the black hole. Concretely, in this paper, we study extensions of EMDs (and observables computed from them) based on \emph{Padé approximants} (see Appendix~\ref{App:DefPadeApproximant} for the definition) of the metric deformations (or suitable functions thereof). The order of the Padé approximant determines how many of the initial Taylor series coefficients are retained and thus consider the physical input to the effective description of the black hole. The infinitely many remaining coefficients can be understood to be replaced by functions of the latter in such a way, to allow applicability of the EMD to a larger region of space-time. We argue in particular, that this approximation can be implemented in a consistent way directly at the level of observables, thus allowing to efficiently compute their approximations.\\

We showcase this approach explicitly for the black hole shadow, by studying the leading quantum correction of the classical result (assuming that it can be defined in an analytic fashion). This correction is a function of finitely many coefficients of the metric deformation, depending on the order of the applied Padé approximant. By increasing this order in a particular fashion, we uncover a pattern in the way these coefficients contribute to the black hole shadow, which allows us to conjecture its asymptotic limit. Moreover, we can (at least formally) resum this limit in terms of the (full) metric deformation function. We therefore find the effective black hole shadow, quantum corrected to leading order, to be proportional to the (full) metric deformation function evaluated at the physical distance of the classical photon sphere. \\

Our result for the shadow takes into account the fully deformed metric function (\emph{i.e.} it does no longer depend on any order of the Padé approximant), however, only takes into account leading order quantum corrections. While in intermediate steps we have counted the order of the latter through the deformation of the position of the event horizon, we have checked that the final result does not depend on the exact choice of this small parameter. Furthermore, although intermediate steps of our derivation assume that the shadow is an analytic function of this parameter, we have verified in explicit examples that the final result holds to good accuracy also in examples where this is not the case: indeed, we have verified and checked our approach and our result for the black hole shadow in a number of examples of quantum deformed, spherically symmetric and static black holes, which have been proposed in the literature in recent years: these include black hole geometries proposed by Hayward~\cite{Hayward_2006}, Simpson and Visser~\cite{Simpson:2018tsi,Simpson:2019mud} and Dymnikova~\cite{Dymnikova:1992ux}, in which cases the quantum deformations are encoded in a single deformation function. We have also studied the metric proposed in \cite{Garcia95} (stemming from an axion-dilaton system) that is characterised by two deformation functions. For all these space-times we demonstrate how Padé approximants can be used to efficiently approximate the metric deformations in a way to accurately capture the leading quantum deformations of physical observables. We in particular showcase the efficiency of the effective result for the black hole shadow, which captures the numerical results (of the fully deformed space-time) with an error of $<1\%$ in the appropriate parameter space. This is much less than the current experimental error of the shadow of the black hole M$87^*$, which is around $17\%$ (see \cite{EventHorizonTelescope:2021dqv}).\\

These results and checks show that the EMDs developed in previous works \cite{Binetti:2022xdi,DelPiano:2023fiw,DelPiano:2024gvw} can be extended to regions of space-time that are further away from the black hole horizon. Moreover, they can be used to calculate effective black hole observables in a universal and model independent fashion. We therefore expect in the future numerous applications of our approach not only in the theoretical study of black holes and their universal properties, but also in experimental studies: indeed, our approach makes it possible to compare measurements of different observables and from different black holes in an unambiguous way that does not require to resort to specific models. Indeed, since our deformation functions are formulated in terms physical quantities, they become directly comparable across different situations.\\

This paper is organised as follows: Section~\ref{Sect:ReviewMetric} provides a review of the effective metric approach introduced in \cite{Binetti:2022xdi,DelPiano:2023fiw,DelPiano:2024gvw} and introduces the black hole shadow as a physical observable. Section~\ref{Sect:PadeApproximant} introduces Padé approximants as a way to extend EMDs to regions outside of the immediate vicinity of the black hole horizon. We showcase how these approximations can be made consistently at the level of the metric deformations and physical observables, notably the proper distance and the black hole shadow. Notably, we provide an effective expression for the black hole shadow, quantum corrected to leading order. Section~\ref{subs example models} illustrates the approach at 5 different examples (\emph{i.e.} models of quantum corrected black hole geometries proposed in the literature) and in particular exhibits our effective results for the shadow. Finally, Section~\ref{Sect:Conclusions} contains our conclusions. This paper is supplemented by two appendices: Appendix~\ref{Sect:Convergence} contains calculational details on the radius of convergence of the Taylor series of the metric deformation functions for some of the concrete black hole models, while Appendix~\ref{App:DefPadeApproximant} contains the mathematical definition and further information of Padé approximants.

\section{Effective Metric Approach and Approximations}\label{Sect:ReviewMetric} 
\subsection{Metric and Proper Distance}
Our starting point is an effective metric description (EMD) (see \cite{DelPiano:2024gvw}) of a spherically symmetric and static black hole, based on the proper distance to the black hole horizon. Indeed, the infinitesimal line element (outside the event horizon) is given by 
\begin{equation}\label{MetricGeneral}
\dd s^{2} = - h(\co) \dd t^{2} + f(\co)^{-1} \dd \co^{2} + \co^{2} \dd \Omega_{2}^2 \ ,
\end{equation}
where $\dd \Omega_2^2 :=  \dd \theta^2 +  \sin^2 \theta \, \dd \varphi^2$ and we parametrise the metric functions in the following form
\begin{equation}
h(\co) = 1 - \frac{\Psi(\rho(\co))}{\co} \qq{and} f(\co) = 1 - \frac{\Phi (\rho(\co))}{\co}  \ .\label{hfgen}
\end{equation}
Here $\Phi$ and $\Psi$ are \emph{a priori} general functions of the distance $\rho$ to the event horizon. The latter satisfies the differential equation
\begin{align}
&\dv{\rho}{\co}=\frac{1}{\sqrt{f(\co)}} \qq{and} \rho(\co=\zh)=0 \ ,\label{DiffEqDistance}
\end{align}
where $\zh$ is the position of the horizon and we assume $\rho\geq 0$ and $r\geq \zh$. Following the notation of \cite{DelPiano:2023fiw}, we assume that the functions $\Psi$ and $\Phi$ can be expanded in the following form\footnote{For later convenience (notably for homogeneity of the functions $\Psi$ and $\Phi$), the coefficients $\{\theta_{2n}\}_{n\in\mathbb{N}}$ and $\{\xi_{2n}\}_{n\in\mathbb{N}}$ have been rescaled with a factor of $2\ma $ relative to the definition in \cite{DelPiano:2023fiw}.}
\begin{equation}
\Psi(\rho)=\zh+ 2\ma\sum_{n=1}^\infty \theta_{2n}\,\rho^{2n}  \qq{and} \Phi(\rho)=\zh+2\ma\sum_{n=1}^\infty \xi_{2n}\,\rho^{2n}\ ,
\label{SeriesExpansionMetric}
\end{equation}
where $\ma$ is the ADM mass (in units of the Planck mass) measured by an observer at space-like infinity. Furthermore, we consider the coefficients  $\{\theta_{2n}\}_{n\in\mathbb{N}}$ and $\{\xi_{2n}\}_{n\in\mathbb{N}}$ (along with $\zh$) the effective parameters describing the black hole under consideration. For compactness of the notation, we shall also sometimes use the notation $\xi_0=\frac{\zh}{2\ma}=\theta_0$ and implicitly understand $\xi_{2n+1}=\theta_{2n+1}=0$ $\forall n\geq 0$.

The differential equation (\ref{DiffEqDistance}) can be solved in terms of the following series expansion \cite{DelPiano:2023fiw,DelPiano:2024gvw} 
\begin{align}
\co=\zh+\sum_{n=1}^\infty a_{2n}\,\rho^{2n}\,,\label{Solz}
\end{align}
where (iteratively) the coefficients $a_{2n}$ have been determined in \cite{DelPiano:2023fiw} as functions of the $\{\xi_{2n}\}_{n\in\mathbb{N}}$: 
\begin{align}
    a_2 &= \frac{1+ \varpi}{8 \zh} \hspace{1cm}\text{with}\hspace{1cm} \varpi = \sqrt{1-32 \ma \zh \xi_2} \ ,\nonumber\\
        a_p&=\frac{1}{1-4\,p\,\zh a_2}\left[2\,\ma\,\xi_p+\zh\sum_{n=3}^{p-1}(p-n+2)\,n\,a_n\,a_{p-n+2}+\sum_{n=2}^{p-2}\sum_{m=2}^n(n-m+2)\,m\,a_{p-n}\,a_m\,a_{n-m+2}\right]\qq{for} p\geq 3 \ .\label{a2expr}
    \end{align}
Notice that since $\xi_{2n+1}=0$ also $a_{2n+1}=0$ $\forall n\geq 0$. By series reversion, knowledge of the $\{a_{2n}\}_{n\in\mathbb{N}}$ allows to write
\begin{align}
\rho=\sum_{n=1}^\infty b_{2n-1}\,(\co-\zh)^{\frac{2n-1}{2}}\,,\label{RhoSolution}
\end{align}
with the coefficients $\{b_{2n-1}\}_{n\in\mathbb{N}}$ explicitly given in \cite{DelPiano:2023fiw}, whose precise form shall not be of relevance in the following. Finally, for further convenience, we introduce two types of series expansions of the metric functions, which (assuming their convergence) are locally equivalent: first we introduce an expansion in terms of the radial coordinate
\begin{align}
&h=\sum_{n=1}^\infty \frac{\hn{n}}{n!}\,(\co-\zh)^n&&\text{and} &&f=\sum_{n=1}^\infty \frac{\fn{n}}{n!}\,(\co-\zh)^n\,.\label{fhnExpansion}
\end{align}
Here we consider the expansion coefficients $\{\hn{n}\}_{n\in\mathbb{N}}$ and $\{\fn{n}\}_{n\in\mathbb{N}}$ as functions of $\{\xi_{2n}\}_{n\in\mathbb{N}}$ and $\{\theta_{2n}\}_{n\in\mathbb{N}}$, for example 
\begin{align}
&\hn{1}=\frac{1+\varpi-16\ma \zh \theta_2}{\zh(1+\varpi)}\,,\nonumber\\
&\hn{2}=\frac{-2048 \theta_2  M^2 \xi_4  \zh ^4+8 M \zh  (\varpi +1) \left(\theta_2  (\varpi +1) (3 \varpi +1)-16
   \theta_4  \zh ^2 (2 \varpi +1)\right)-(\varpi +1)^3 (2 \varpi +1)}{\zh ^2 (\varpi +1)^3 (2 \varpi +1)}\,.
\end{align}
Secondly, we write
\begin{align}
&h(\rho)=1-2\ma\sum_{p=0}^\infty \left(\sum_{n=0}^p\theta_{2p-2n}\,s_{2n}\right) \rho^{2p}\,,&&\text{and} &&f(\rho)=1-2\ma\sum_{p=0}^\infty \left(\sum_{n=0}^p\xi_{2p-2n}\,s_{2n}\right) \rho^{2p}\,,\label{fhRho}
\end{align}
where have introduced the coefficients $\{s_{2n}\}_{n\in\mathbb{N}}$ as functions of the $\{a_{2n}\}_{n\in\mathbb{N}}$ via 
\begin{align}
&\frac{1}{\co}=\sum_{\ell=0}^\infty\,s_{2\ell}\,\rho^{2\ell}&&\text{with} &&s_0=\dfrac{1}{\zh} \qq{} \text{and} \qq{} s_{2p}=\displaystyle -\dfrac{1}{\zh}\sum_{\ell=0}^{p-1}a_{2p-2\ell}s_{2\ell}\hspace{0.5cm}\forall p>0\,.\label{SCoeffs}
\end{align}
We stress that both expansions (\ref{fhnExpansion}) and (\ref{fhRho}) in general have finite radius of convergence in a region close to the event horizon. Thus, while these expansions are crucial in (locally) solving (\ref{DiffEqDistance}) in a self-consistent fashion, and thus determining explicitly the geometry, they are in general not viable for computing observables at a distance from the black hole. Furthermore, even close to the horizon, convergence of (\ref{SeriesExpansionMetric}) (or (\ref{fhnExpansion})) may be slow, thus making them impractical for concrete computations.\footnote{We note, however, that (\ref{SeriesExpansionMetric}), (\ref{fhnExpansion}) and (\ref{fhRho}) allow to compute observables that only depend on the (very near-)horizon geometry in a very efficient manner, such as the Hawking temperature of the black hole \cite{DelPiano:2023fiw,DelPiano:2024gvw}.}  The main result of this work is therefore to devise a strategy for approximations of the metric that circumvent these problems: in the following Subsection~\ref{Sect:PhotonRadius} we shall briefly introduce an observable, in the form of the photon radius (which is experimentally related to the black hole shadow), which requires a description of the geometry at a distance of the black hole horizon. Furthermore, in Section~\ref{Sect:PadeApproximant} we shall introduce the Padé approximant as a means to (approximately) calculate this observable in the effective framework.

Before continuing, we would like to make a further technical remark concerning the series coefficients in (\ref{SeriesExpansionMetric}) mentioned above: indeed, since $\Psi$ and $\Phi$ encode (quantum) corrections, their effect is in a certain sense understood to be small. In fact, in many models (see Section~\ref{subs example models} below for examples) there is a dedicated small parameter (denoted $\eta$ in Section~\ref{subs example models}) that governs the size of the deformation and allows to (better) organise approximations accordingly. 
%In the case that such a dedicated parameter does not exist (or is not immediately apparent), we can imagine to organise the expansion in inverse powers of $\ma$ (since the black hole mass is a scale which we have assumed to exist also in the absence of the deformation functions $\Psi$ and $\Phi$), which is very small at least for astrophysical black holes. We shall then assume that the parameters $\{\theta_{2n},\xi_{2n}\}_{n\in\mathbb{N}}$ (as well as $\zh$) are functions of $\ma$. 
In order to make contact with the examples discussed in Section~\ref{subs example models}, we shall also define
\begin{align}
&\theta_{2n}=:\frac{\mathfrak{t}_{2n}}{(2\ma)^{2n}}\,, &&\text{and} &&\xi_{2n}=:\frac{\mathfrak{x}_{2n}}{(2\ma)^{2n}}\,,&&\text{and} &&\zh=:2\ma\left(1+\mathfrak{c}\right)\,,\label{RescaledMetricFunctions}
\end{align}
and assume that $\{\mathfrak{t}_{2n}\,,\mathfrak{x}_{2n}\}$ and $\mathfrak{c}$ are small deformations. Concretely, in order to illustrate some of our conceptual points by an example, we shall consider
\begin{align}
&\mathfrak{c}\ll 1&&\text{and}&&\frac{\mathfrak{t}_{2n}}{\mathfrak{c}}\gg \mathfrak{c}\,,&&\text{and} &&\frac{\mathfrak{x}_{2n}}{\mathfrak{c}}\gg \mathfrak{c}\,,&&\forall n\geq 2\,.\label{IllumScaling}
\end{align}
Assuming observables to be analytic functions in $\mathfrak{c}$, we present our results as expansions thereof, focusing mostly on the leading correction $\mathcal{O}(\mathfrak{c})$. We shall consider this, the \emph{leading quantum correction}.%Furthermore, in order to make close contact with some of the examples discussed in Section~\ref{subs example models}, we shall write 
%\begin{align}
%&\mathfrak{c}=-\frac{\epsilon}{1+\epsilon}\,,&&\text{with} &&\epsilon\ll 1\,,
%\end{align}
%and present our results as (analytic) expansions in powers of $\epsilon$. 

%%%%%%%%%%%%%%%%%%%%%%%%%%%%%%%%%%
\subsection{Radius of the Photon Sphere and Black Hole Shadow}\label{Sect:PhotonRadius}
Null-geodesics that describe the motion of massless particles in the geometry (\ref{MetricGeneral}) satisfy the following equation 
\begin{align}\label{effradeq}
&    \left(\dv{r}{\varphi} \right)^2 = r^2 f(r)\left( \frac{r^2}{b^2 h(r)}-1 \right) \ ,&&\text{with} &&b := L/E\,,
\end{align}
where the impact parameter $b$ is the quotient of the angular momentum $L$ and the energy $E$, which are conserved along the geodesic. Equation \eqref{effradeq} can be read as a one-dimensional energy conservation law \cite{Perlick:2021aok,Carroll:2004st}  in the sense $(\dd r/\dd \varphi)^2 + V_\text{eff}=0$, with the effective potential $ V_\text{eff}$ given by the right hand side in (\ref{effradeq}). The particle follows a circular orbit (\emph{i.e.} with $r=\rps=$const.) for $(\rps,b)$ that satisfy 
\begin{align}
 &V_\text{eff} = 0 \qq{and} \dv{V_\text{eff}}{r}=0 &&\text{such that} &&\dv{r}{\varphi}\bigg|_{r=\rps} = \dv[2]{r}{\varphi}\bigg|_{r=\rps} = 0\,.\label{PhotonRadius}
\end{align}

\begin{figure}[t]
    \centering
\tikzset{every picture/.style={line width=0.75pt}} %set default line width to 0.75pt        
\resizebox{0.65\columnwidth}{!}{
\begin{tikzpicture}[x=0.75pt,y=0.75pt,yscale=-1,xscale=1]

%Shape: Circle [id:dp7781590149480428] 
\draw  [draw opacity=0][fill={rgb, 255:red, 0; green, 0; blue, 0 }  ,fill opacity=1 ] (86,140) .. controls (86,121.22) and (101.22,106) .. (120,106) .. controls (138.78,106) and (154,121.22) .. (154,140) .. controls (154,158.78) and (138.78,174) .. (120,174) .. controls (101.22,174) and (86,158.78) .. (86,140) -- cycle ;
%Shape: Circle [id:dp5265833196925818] 
\draw  [color={rgb, 255:red, 208; green, 2; blue, 27 }  ,draw opacity=1 ] (65,140) .. controls (65,109.62) and (89.62,85) .. (120,85) .. controls (150.38,85) and (175,109.62) .. (175,140) .. controls (175,170.38) and (150.38,195) .. (120,195) .. controls (89.62,195) and (65,170.38) .. (65,140) -- cycle ;
%Straight Lines [id:da814957216437622] 
\draw [color={rgb, 255:red, 255; green, 255; blue, 255 }  ,draw opacity=1 ]   (120,140) -- (135.07,167.09) ;
\draw [shift={(136.53,169.71)}, rotate = 240.91] [fill={rgb, 255:red, 255; green, 255; blue, 255 }  ,fill opacity=1 ][line width=0.08]  [draw opacity=0] (5.36,-2.57) -- (0,0) -- (5.36,2.57) -- cycle    ;
%Straight Lines [id:da32988812080396035] 
\draw [color={rgb, 255:red, 208; green, 2; blue, 27 }  ,draw opacity=1 ]   (120,140) -- (79.77,172.95) ;
\draw [shift={(77.45,174.85)}, rotate = 320.68] [fill={rgb, 255:red, 208; green, 2; blue, 27 }  ,fill opacity=1 ][line width=0.08]  [draw opacity=0] (5.36,-2.57) -- (0,0) -- (5.36,2.57) -- cycle    ;
%Straight Lines [id:da12730790496185818] 
\draw [color={rgb, 255:red, 74; green, 144; blue, 226 }  ,draw opacity=1 ]   (120,140) -- (100.82,72.88) ;
\draw [shift={(100,70)}, rotate = 74.05] [fill={rgb, 255:red, 74; green, 144; blue, 226 }  ,fill opacity=1 ][line width=0.08]  [draw opacity=0] (5.36,-2.57) -- (0,0) -- (5.36,2.57) -- cycle    ;
%Straight Lines [id:da9661540096312462] 
\draw [color={rgb, 255:red, 74; green, 144; blue, 226 }  ,draw opacity=1 ] [dash pattern={on 4.5pt off 4.5pt}]  (60,10) -- (362.82,140) ;
%Straight Lines [id:da35255205303867587] 
\draw  [dash pattern={on 0.84pt off 2.51pt}]  (120,140) -- (362.82,140) ;
%Curve Lines [id:da20228783455576882] 
\draw [color={rgb, 255:red, 150; green, 144; blue, 226 }  ,draw opacity=1]   (100,70) .. controls (181.33,54.91) and (350.33,135.92) .. (362.82,140) ;
%\draw [shift={(234.78,89.95)}, rotate = 18.14] [fill={rgb, 255:red, 74; green, 144; blue, 226 }  ,fill opacity=1 ][line width=0.08]  [draw opacity=0] (5.36,-2.57) -- (0,0) -- (5.36,2.57) -- (3.56,0) -- cycle    ;
%\draw [shift={(100,70)}, rotate = 347.01] [fill={rgb, 255:red, 74; green, 144; blue, 226 }  ,fill opacity=1 ][line width=0.08]  [draw opacity=0] (5.36,-2.57) -- (0,0) -- (5.36,2.57) -- (3.56,0) -- cycle    ;
%Curve Lines [id:da16403243585033245] 
\draw    (363.82,144.83) .. controls (361.4,143.93) and (361.09,136.67) .. (363.82,135.67) ;
%Curve Lines [id:da6698787608455452] 
\draw    (370,140.69) .. controls (367.84,139.79) and (364.39,139.79) .. (362.52,132.39) ;
%Shape: Boxed Bezier Curve [id:dp7285286448963018] 
\draw    (370,140.69) .. controls (367.84,141.59) and (364.39,141.59) .. (362.52,148.99) ;
%Shape: Boxed Bezier Curve [id:dp8646112305862614] 
\draw [fill={rgb, 255:red, 0; green, 0; blue, 0 }  ,fill opacity=1 ]   (362.04,143.04) .. controls (363.32,142.46) and (363.49,137.79) .. (362.04,137.14) ;

%Straight Lines [id:da2292646338254718] 
\draw [color={rgb, 255:red, 74; green, 144; blue, 226 }  ,draw opacity=1 ] [dash pattern={on 4.5pt off 4.5pt}]  (280,140) -- (362.82,140) ;
%Curve Lines [id:da2445721906224072] 
\draw [color={rgb, 255:red, 150; green, 144; blue, 226 }  ,draw opacity=1 ]   (10.61,177) .. controls (16.82,148.14) and (38.65,85.85) .. (100,70) ;
%\draw [shift={(39.01,112.33)}, rotate = 307.61] [fill={rgb, 255:red, 74; green, 144; blue, 226 }  ,fill opacity=1 ][line width=0.08]  [draw opacity=0] (5.36,-2.57) -- (0,0) -- (5.36,2.57) -- (3.56,0) -- cycle    ;
%\draw [shift={(10,180)}, rotate = 280.96] [fill={rgb, 255:red, 74; green, 144; blue, 226 }  ,fill opacity=1 ][line width=0.08]  [draw opacity=0] (5.36,-2.57) -- (0,0) -- (5.36,2.57) -- (3.56,0) -- cycle    ;
%Straight Lines [id:da9829674300211868] 
\draw [color={rgb, 255:red, 74; green, 144; blue, 226 }  ,draw opacity=1 ] [dash pattern={on 4.5pt off 4.5pt}]  (280,140) -- (280,106.11) ;
%Shape: Arc [id:dp47637747028277144] 
\draw  [draw opacity=0][dash pattern={on 1.69pt off 2.76pt}][line width=1.5]  (280,140) .. controls (280,139.94) and (280,139.87) .. (280,139.81) .. controls (280,128.48) and (282.93,117.83) .. (288.08,108.59) -- (344.21,139.81) -- cycle ; \draw  [color={rgb, 255:red, 74; green, 144; blue, 226 }  ,draw opacity=1 ][dash pattern={on 1.69pt off 2.76pt}][line width=1.5]  (280,140) .. controls (280,139.94) and (280,139.87) .. (280,139.81) .. controls (280,128.48) and (282.93,117.83) .. (288.08,108.59) ;  
%Straight Lines [id:da47512989450001997] 
\draw [color={rgb, 255:red, 74; green, 144; blue, 226 }  ,draw opacity=1 ]   (120,140) -- (156.4,54.62) ;
\draw [shift={(157.57,51.86)}, rotate = 113.09] [fill={rgb, 255:red, 74; green, 144; blue, 226 }  ,fill opacity=1 ][line width=0.08]  [draw opacity=0] (7.14,-3.43) -- (0,0) -- (7.14,3.43) -- (4.74,0) -- cycle    ;

% Text Node
\draw (131,142.4) node [anchor=north west][inner sep=0.75pt]  [font=\scriptsize,color={rgb, 255:red, 255; green, 255; blue, 255 }  ,opacity=1 ]  {$\hr$};
% Text Node
\draw (85,168) node [anchor=north west][inner sep=0.75pt]  [font=\scriptsize,color={rgb, 255:red, 208; green, 2; blue, 27 }  ,opacity=1 ]  {$\rps$};
% Text Node
\draw (285.83,116.17) node [anchor=north west][inner sep=0.75pt]  [color={rgb, 255:red, 74; green, 144; blue, 226 }  ,opacity=1 ]  {$\alpha $};
% Text Node
\draw (298.38,144.76) node [anchor=north west][inner sep=0.75pt]  [font=\scriptsize,color={rgb, 255:red, 74; green, 144; blue, 226 }  ,opacity=1 ]  {$\sqrt{g_{rr}} \dd r$};
% Text Node
\draw (226.06,114.63) node [anchor=north west][inner sep=0.75pt]  [font=\scriptsize,color={rgb, 255:red, 74; green, 144; blue, 226 }  ,opacity=1 ]  {$\sqrt{g_{\varphi \varphi }} \dd \varphi $};
% Text Node
\draw (106.4,72) node [anchor=north west][inner sep=0.75pt]  [font=\scriptsize,color={rgb, 255:red, 74; green, 144; blue, 226 }  ,opacity=1 ]  {$\rmin$};
% Text Node
\draw (149.83,74.46) node [anchor=north west][inner sep=0.75pt]  [font=\scriptsize,color={rgb, 255:red, 74; green, 144; blue, 226 }  ,opacity=1 ]  {$b$};
% Text Node
\draw (358.97,149.89) node [anchor=north west][inner sep=0.75pt]  [font=\scriptsize,color={rgb, 255:red, 74; green, 144; blue, 226 }  ,opacity=1 ]  {$r_\text{O}$};

\end{tikzpicture}
}

\caption{Observer-dependent inclination angle $\alpha$ and the minimal radial distance $\rmin$ for a photon on an open trajectory around the black hole. The impact parameter $b$ is measured by an observer at the (large) distance $\robs$. The red circle $\rps$ represents the photon ring (\emph{i.e.} a circular orbit for massless particles around the black hole).}
    \label{fig:bhshadow}
\end{figure}
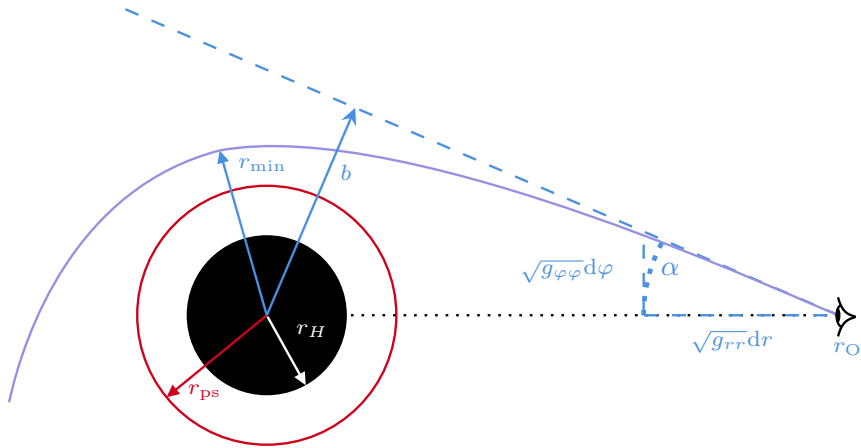

It is convenient to re-formulate the problem in terms of the minimal radius distance $\rmin$ of a massless particle on a general open trajectory  \cite{Perlick:2021aok,Carroll:2004st,misner1973gravitation} (see Figure~\ref{fig:bhshadow}), such that (\ref{effradeq}) becomes~\cite{Bozza:2002zj}
\begin{align}
&\left(\dv{r}{\varphi}\right)^2 =  r^2 f(r)\left( \frac{U(\rmin)^2}{U(r)^2} - 1 \right) \ ,&&\text{with} &&U(r)^2=\frac{h(r)}{r^2}\,,\label{PhotonPotentialU}
\end{align}
where the new effective potential satisfies $U(\rmin)=b^{-1}$. Furthermore, the inclination angle $\alpha$ measured by a static observer located at a (large) radial distance $\robs$ is \cite{Synge1966,Perlick:2015vta,Perlick:2021aok}
\begin{align}
&\cot \alpha = \frac{1}{\sqrt{r^2f(r)}} \left. \dv{r}{\varphi} \right|_{r=\robs} &&\text{such that} &&  \sin^2 \alpha = \frac{U(\robs)^2}{U(\rmin)^2}\ ,
\end{align}
from which we can obtain the angular aperture of the black hole shadow as an observer-dependent quantity
\begin{equation}\label{bphdef}
    \sin^2 \alpha_\text{sh} = U(\robs)^2 \bsh^2 \qq{with} \bsh := U(\rps)^{-1} \ .
\end{equation}
Here $\bsh$ is usually called the \emph{radius of the black hole shadow} \cite{Perlick:2021aok} and, as before, $\rps$ is the radius of the photon sphere, which is computed from (\ref{PhotonRadius}). Using the potential in (\ref{PhotonPotentialU}), the latter condition becomes
\begin{equation}\label{rpheq}
    \left.\dv{U^2(r)}{r}\right|_{r=\rps}=0 \implies  \left\{\begin{array}{l}\rps h^\prime(\rps) = 2 h(\rps) \hspace{0.2cm}\text{or equivalently} \\[4pt] 3\Psi(\rho_{\text{ps}})=\rps\,\left(2+\dv{\rho}{\co}\,\dv{\Psi}{\rho}\big|_{\co=\rps}\right)\,.
\end{array}\right.
\end{equation}
In principle, there may be more than one solution to this equation, allowing for the presence of several photon rings, some of which are stable (around which photons oscillate) and some unstable (towards which photons spiral) \cite{Gan:2021pwu}. However, following \cite{Cunha:2020azh}, we expect at least one local maximum for the effective potential $U$ outside of the event horizon.

As we shall see explicitly in several examples in Section~\ref{subs example models}, the radius of the photon sphere $\rps$ is not only larger than the radius of the event horizon $\zh$, but also such that the series expansions (\ref{DiffEqDistance}) or the expansion of the distance in (\ref{RhoSolution}) are not necessarily convergent, or at least converge very slowly. Thus, they cannot be used directly to find the solution (\ref{rpheq}). In the following Section we shall therefore present approximations of the metric that allow to (approximately) compute $\rps$, the impact parameter $b$ and ultimately the shadow $\bsh$.

%%%%%%%%%%%%%%%%%
\section{Padé Approximation}\label{Sect:PadeApproximant}
In this Section we use Padé approximants (see Appendix~\ref{App:DefPadeApproximant} for the definition) to find approximations to the effective metric description and physical observables in a region outside of the event horizon, notably the radius of the photon ring discussed in Section~\ref{Sect:PhotonRadius}. We shall introduce the concept of an order of these approximations through the number of expansion coefficients $\{\xi_{2n},\theta_{2n}\}_{n\in\mathbb{N}}$ that we consider to be genuine and independent, which is linked to the order of the Padé approximant.

%%%%%%%%%%%%%%%%%%%%%%%%%%%%%%%%%%
\subsection{Padé Approximant of the Metric and Horizon Distance}\label{Sect:PadeMetric}
While the series expansions (\ref{SeriesExpansionMetric}) for the metric deformation functions, may exist close to the horizon of the black hole, they are generally expected to have finite radius of convergence. Thus (\ref{SeriesExpansionMetric}) can in general not be directly used to describe the space time at an arbitrary finite distance away from the event horizon. We therefore propose to \emph{approximate} the metric deformations $\Psi$ and $\Phi$ by new functions, which are defined for (general) $\rho>0$, while still representing $\Psi$ and $\Phi$ well, close to the horizon. Concretely, we propose to approximate the deformation functions by their Padé approximants, as defined in Appendix~\ref{App:DefPadeApproximant}
\begin{align}
&\Psi(\rho)\sim\zh+\Psi_{N,M}(\rho)\,,&&\text{and} && \Phi(\rho)\sim\zh+ \Phi_{N,M}(\rho)\,.\label{PadeApproximation}
\end{align} 
In terms of the formalism developed in \cite{DelPiano:2023fiw,DelPiano:2024gvw}, this approximation implies that the coefficients $\{\theta_{2n}\}_{2n>M+N}$ and $\{\xi_{2n}\}_{2n>M+N}$ are replaced
\begin{align}
&\theta_{2n}\longrightarrow \widetilde{\theta}_{2n}(\theta_2,\ldots,\theta_{2L})&&\text{and}&&\xi_{2n}\longrightarrow \widetilde{\xi}_{2n}(\xi_2,\ldots,\xi_{2L})\,,&&\forall 2n>L:=\left\lfloor\frac{M+N}{2}\right\rfloor\,,\label{ReplacementCoefficients}
\end{align}
by functions of $\theta_2,\ldots,\theta_{2L}$ and $\xi_2,\ldots,\xi_{2L}$ respectively. In the following, we shall consider these latter parameters as physically significant and interpret (\ref{ReplacementCoefficients}) (stemming from (\ref{PadeApproximation})) as an approximation.

Since, according to (\ref{a2expr}), $a_{2n}(\xi_2,\ldots,\xi_{2n})$ is independent of $\xi_{2m}$ for $m>n$, the approximation (\ref{PadeApproximation}) also leads to a modified solution $\widetilde{\rho}$ of (\ref{DiffEqDistance}) for the physical distance, which can be written in the form (\ref{RhoSolution})
\begin{align}
&\widetilde{\rho}(\co)=\sum_{k=1}^{L}b_{2k-1}\,(\co-\zh)^{\frac{2k-1}{2}}+\sum_{k=L+1}^{\infty}\widetilde{b}_{2k-1}\,(\co-\zh)^{\frac{2k-1}{2}}\,,\label{ExactApproximation}
\end{align}
where we have also used the fact that the $b_{2n-1}(a_2,\ldots,a_{2n})$ in (\ref{RhoSolution}) are independent of $a_{2m}$ for $m>n$. The $\widetilde{b}_{2k-1}(\xi_2,\ldots, \xi_{2L})$ are implicit functions of the remaining parameters $\{\xi_2,\ldots,\xi_{2L}\}$, which can be obtained through series reversion of 
\begin{align}
\co=\zh+\sum_{k=1}^{L} a_{2k}\,\rho^{2k}+\sum_{k=L+1}^{\infty} \widetilde{a}_{2k}\,\rho^{2k}\,,
\end{align}
with the coefficients
\begin{align}
\widetilde{a}_p&=\frac{1}{1-4\,p\,\zh a_2}\left[2\,\ma\,\widetilde{\xi}_p+\zh\sum_{n=3}^{p-1}(p-n+2)\,n\,\widetilde{a}_n\,\widetilde{a}_{p-n+2}+\sum_{n=2}^{p-2}\sum_{m=2}^n(n-m+2)\,m\,\widetilde{a}_{p-n}\,\widetilde{a}_m\,\widetilde{a}_{n-m+2}\right]\qq{for} p> 2L \ .\label{ModCoeffsA}
\end{align}
Here we understand $\widetilde{a}_{k}=a_{k}$ for $k\leq 2L$ with $a_{k}$ given in (\ref{a2expr}). We remark that $\widetilde{\rho}$ is a solution of 
\begin{align}
\dv{\widetilde{\rho}}{\co}=\frac{1}{\sqrt{\widetilde{f}(\co)}} \qq{with} \widetilde{f}(\co)=1-\frac{\zh+\Phi_{N,M}}{\co}\ .\label{ReducedDiff}
\end{align}
Notice furthermore that $a_{2k}\neq \widetilde{a}_{2k}$ for $k>L$ due to the fact that (in general) $\widetilde{\xi}_{2k'}\neq \xi_{2k'}$ for $L<k'\leq k$. If the differences  $|a_{2k}-\widetilde{a}_{2k}|$ are small (which certainly depends on the $\{\xi_{2n}\}_{n\in\mathbb{N}}$ and which we shall analyse in a number of examples in Section~\ref{subs example models}), (\ref{ExactApproximation}) constitutes a good approximation for the exact physical distance (\ref{RhoSolution}). However, even if $\widetilde{\rho}$ constitutes a viable approximation, the form (\ref{ExactApproximation}) may not necessarily have better convergence properties than (\ref{fhnExpansion}). For this reason, we shall introduce yet another approximation 
\begin{align}
\widehat{\rho}:=\rho_{N,M}(\sqrt{\co-\zh})=\sum_{k=1}^{L}b_{2k-1}\,(\co-\zh)^{\frac{2k-1}{2}}+\sum_{k=L+1}^{\infty}\widehat{b}_{2k-1}\,(\co-\zh)^{\frac{2k-1}{2}}\,,\label{PadeDistance}
\end{align}
namely the Padé approximant of the distance function, which is implicitly determined by the coefficients $\{b_1,\ldots,b_{2L-1}\}$ that are given in terms of $\{a_2,\ldots,a_{2L}\}$ and thus in terms of $\{\xi_2,\ldots,\xi_{2L}\}$. Notice that $\widehat{\rho}$ only carries information about the coefficients $\xi_{2,\ldots,2L}$ but not $\xi_{2k}$ for $k>L$. In this sense, it is an approximation to the same order as (\ref{ExactApproximation}), except that in general $\widetilde{b}_{2k-1}\neq \widehat{b}_{2k-1}$ for $k>L$. Furthermore, we stress that $\widehat{\rho}$ is a solution of (\ref{DiffEqDistance}) and of (\ref{ReducedDiff}) only up to order $(\co-\zh)^{\frac{2L-1}{2}}$.

%%%%%%%%%%%%%%%%%%%%%%%%%%%%%%%%%%%%%%%%

\begin{figure}[t]
    \centering
    \includegraphics[width=0.55\textwidth]{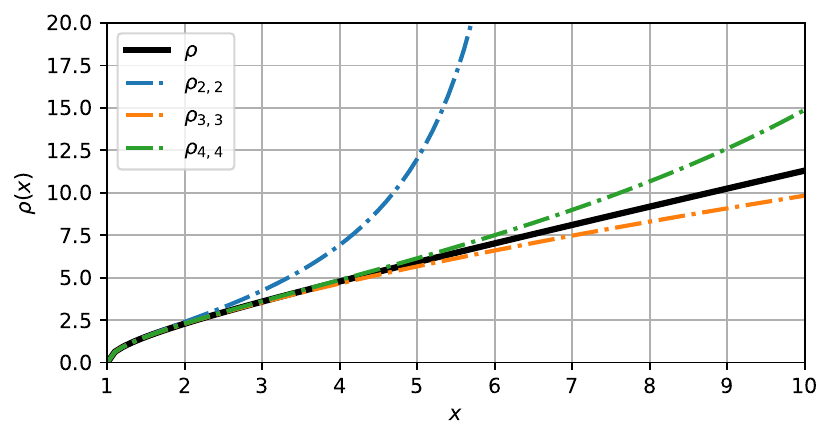}
    \caption{Different Padé approximants of the proper distance to the horizon in the case of the Schwarzschild black hole (with $\Psi=\Phi=2\ma$.)}
    \label{fig:disSchwarz}
\end{figure}

To better illustrate the above discussion, we explicitly consider the example $N=M=2$, such that
\begin{align}
&&\Psi_{2,2}(\rho)=2\ma\,\frac{\left|\begin{array}{cccc}0 & \theta_2 & 0\\\theta_2 & 0 & \theta_4 \\ 0 & 0 & \theta_2 \rho^2 \end{array}\right|}{\left|\begin{array}{ccc}0 & \theta_2 & 0 \\ \theta_2 & 0 & \theta_4 \\ \rho^2 & \rho & 1\end{array}\right|}=2\ma\,\frac{\theta_2\rho^2}{1-\frac{\theta_4 \rho^2}{\theta_2}}\,,&&\text{and}&&\Phi_{2,2}(\rho)=2\ma\,\frac{\left|\begin{array}{cccc}0 & \xi_2 & 0\\\xi_2 & 0 & \xi_4 \\ 0 & 0 & \xi_2 \rho^2 \end{array}\right|}{\left|\begin{array}{ccc}0 & \xi_2 & 0 \\ \xi_2 & 0 & \xi_4 \\ \rho^2 & \rho & 1\end{array}\right|}=2\ma\,\frac{\xi_2\rho^2}{1-\frac{\xi_4 \rho^2}{\xi_2}}\,.
\end{align}
In this simple case, the approximation amounts to the replacement (\ref{ReplacementCoefficients}) with
\begin{align}
&\widetilde{\theta}_{2n}=2\ma\,\frac{\theta_4^{n-1}}{\theta_2^{n-2}}\,,
&&\text{and} &&\widetilde{\xi}_{2n}=2\ma\,\frac{\xi_4^{n-1}}{\xi_2^{n-2}}\,,&& \forall n>2\,. \label{exampleP22}
\end{align}
This in particular implies that the series expansions of  $\Psi_{2,2}$ and $\Phi_{2,2}$ have radii of convergence $\frac{\theta_4}{\theta_2} $ and $\frac{\xi_4}{\xi_2}$, respectively.\footnote{The functions (\ref{exampleP22}) have a pole at $\rho=\sqrt{\frac{\theta_2}{\theta_4}}$ and $\rho=\sqrt{\frac{\xi_2}{\xi_4}}$ respectively, but are else well defined for $\rho\geq 0$.} Furthermore, we obtain for the coefficients (\ref{ModCoeffsA})
\begin{align}
&a_2=\frac{1+ \varpi}{8 \zh}\,,&&a_4=-\frac{2\ma\xi_4+\frac{\left(1+\varpi\right)^3}{128 \zh^3}}{1+2\varpi}\,,&&\text{and}&&\widetilde{a}_{2n}=\frac{a_4^{n-1}}{a_2^{n-2}}\hspace{0.5cm}\forall n>2\,.
\end{align}
We furthermore have for (\ref{PadeDistance})
\begin{align}
&\widehat{\rho}\,(\co)=\rho_{2,2}(\co)=\frac{b_1 (\co-\zh)^{1/2}}{1-\frac{b_3}{b_1}\,(\co-\zh)}\,,&&\text{with} &&\begin{array}{l}b_1=\frac{1}{\sqrt{a_2}}\,, \\[4pt] b_3=-\frac{a_4}{2a_2^{5/2}}\ .\end{array}
\end{align}
In order to quantify the difference of $\widehat{\rho}$ and $\rho$ we expand the coefficients $\widehat{b}_{2n-1}$ and $b_{2n-1}$ in powers of $\mathfrak{c}$, assuming for illustrational purposes a scaling of all (initial) coefficients of the form given in (\ref{IllumScaling})
\begin{align}
(2\ma)^{3/2}\,\widehat{b}_5&=\frac{1}{18}+\frac{1}{108}\left(-9+28 \,\frac{\mathfrak{x}_2}{\mathfrak{c}}+192\,\frac{\mathfrak{x}_4}{\mathfrak{c}}\right)\,\mathfrak{c}+\mathcal{O}(\mathfrak{c}^2)\ , \\ 
(2\ma)^{3/2}\,b_5&=-\frac{1}{20}+\frac{1}{5}\left(\frac{3}{8}+\frac{7}{90}\,\frac{\mathfrak{x}_2}{\mathfrak{c}}+\frac{56}{3}\,\frac{\mathfrak{x}_4}{\mathfrak{c}}+64\,\frac{\mathfrak{x}_6}{\mathfrak{c}}\right)\,\mathfrak{c}+\mathcal{O}(\mathfrak{c}^2)\,.
\end{align}
We stress that the differences between $\widehat{b}_5$ and $b_5$, have two conceptually different origins:
\begin{enumerate}
\item[\emph{(i)}] $\rho$ and $\widehat{\rho}$ are defined in two different ways. While this of course impacts the full form of $\widehat{b}_5$, it in particular also concerns the coefficient at order $\mathcal{O}(\mathfrak{c}^{0})$, which is independent of  $\{\mathfrak{t}_{2n}\,,\mathfrak{x}_{2n}\}$: indeed, $\rho$ and $\widehat{\rho}$ are different even in the undeformed (Schwarzschild) black hole, as is illustrated in Figure~\ref{fig:disSchwarz}.
\item[\emph{(ii)}] $\widehat{b}_5$ is obtained through a Padé approximation and therefore only depends on the coefficients $\mathfrak{c}$, $\mathfrak{x}_2$ and $\mathfrak{x}_4$, but not $\mathfrak{x}_6$, unlike $b_5$. Indeed, the coefficient $\widetilde{\mathfrak{x}}_6$ (which enters into $\widehat{b}_5$) is a function of $\mathfrak{x}_2$ and $\mathfrak{x}_4$, which thus contributes also to the numerical differences of the coefficient of order $\mathcal{O}(\mathfrak{c})$. We notice that these modifications in fact only affect terms of order~$\mathcal{O}(\mathfrak{c})$.
\end{enumerate}

As we shall see in more detail below, corrections of type \emph{(ii)} are unavoidable, but become progressively smaller (\emph{i.e.} appear at higher and higher order in the distance expansion) for sufficiently large $N$, $M$. Corrections of the type \emph{(i)} can be avoided for certain observables: indeed, we shall see below, that observables that are rational functions (in a certain variable) in the case of the undeformed black hole have a particular status. For such observables, the Padé approximant for the Schwarzschild black hole becomes exact beyond a certain order, such that corrections of the type \emph{(i)} are absent at this point.

%%%%%%%%%%%%%%%%%%%%%%%%%%%%%%%%%%%%%%%%%%%%%%

%%%%%%%%%%%%%%%%%%%%%%%%%%%%%%%%%%%
\subsection{Padé Approximation for the Photon Radius}\label{Sect:PadePhoton}
We next turn to the computation of the photon ring radius. Using the series expansions (\ref{SeriesExpansionMetric}) and (\ref{Solz}), we can write for the potential in (\ref{PhotonPotentialU})
\begin{align}
U^2&=\sum_{p=1}^\infty \uva{2p}\,\rho^{2p}%=\left(1-\sum_{k=0}^\infty \rho^{2k}\sum_{n=0}^k\theta_{2k-2n}\,s_{2n}\right)\left(\frac{1}{\zh^2}+\sum_{\ell=1}^\infty q_{2\ell}\rho^{2\ell}\right)\nonumber\\
%&=\frac{1}{\zh^2}-\frac{1}{\zh^2}\sum_{p=0}^\infty \rho^{2p}\sum_{n=0}^p\theta_{2p-2n}\,s_{2n}+\sum_{p=1}^\infty q_{2p}\,\rho^{2p}-\sum_{p=1}^\infty \rho^{2p}\sum_{\ell=1}^pq_{2\ell}\sum_{n=0}^{p-\ell} \theta_{2(p-\ell-n)}s_{2n}\nonumber\\
=\sum_{p=1}^\infty\ \rho^{2p}\left[q_{2p}-\frac{2\ma}{\zh^2} \sum_{n=0}^p \theta_{2(p-n)}s_{2n}-2\ma\sum_{\ell=1}^p q_{2\ell} \sum_{n=0}^{p-\ell} \theta_{2(p-\ell-n)} s_{2n}\right]\,,\label{PotentialDistanceExpansion}
\end{align}
where $s_{2n}$ is given in (\ref{SCoeffs}) and 
\begin{align}
q_{2p}=\left\{\begin{array}{lcl}\displaystyle -\frac{4a_2}{\zh^3} & \text{for} & p=1\,,\\ \displaystyle-\frac{1}{2p\zh}\left[\frac{4p a_{2p}}{\zh^2}+\sum_{\ell=1}^{p-1}2(2p-\ell)q_\ell a_{2(p-\ell)}\right] & \text{for}& p>1\,. \end{array}\right.
\end{align}
Concretely, the first few coefficients $\uva{2,4,6}$ of $U$ read
%\begin{subequations}
\begin{align}
\uva{2}&=\frac{1-16\ma\zh \theta_2+\varpi}{8\zh^4}\ ,\\ \uva{4}&=-\frac{(1+\varpi)^2(7+13\varpi)}{128\zh^6(1+2\varpi)}+\frac{3\ma\theta_2(1+\varpi)}{4\zh^5}-2\ma\frac{\xi_4+(1+2\varpi)\theta_4}{\zh^3(1+2\varpi)}\ ,\\
\uva{6}&=-\frac{64\ma^2 \xi_4^2}{(3 \varpi +2) \zh(1+2 \varpi)^2}-\frac{1}{\zh^4} \left(\frac{12\ma^2 \theta_2 \xi_4}{2 \varpi
   +1}+2\ma\zh \left(\theta_6+\frac{\xi_6}{3 \varpi +2}\right)\right)-\frac{3 \ma\theta_2
   (\varpi +1)^2 (9 \varpi +5)}{64 (2 \varpi +1) \zh^7}+ \nonumber\\
    & \quad +\frac{\ma (\varpi +1) \left(6 \theta_4 (3
   \varpi +2) (2 \varpi +1)^2+ \xi_4 (13 \varpi  (6 \varpi +7)+25)\right)}{8 (2 \varpi +1)^2 (3 \varpi +2) \zh^5}+\frac{(\varpi +1)^3 (\varpi  (\varpi  (368 \varpi +655)+386)+75)}{2048 (2 \varpi +1)^2 (3
   \varpi +2) \zh^8}\ .
\end{align}
%\end{subequations}
For finite $p>0$, $\uva{2p}(\zh,\xi_{2,4,\ldots,2p},\theta_{2,4,\ldots,2p})$ only depends on a finite number of the expansion coefficients of the metric deformations. 

Following the example of the approximation of the proper distance in the previous Subsection, we can use (\ref{PotentialDistanceExpansion}) to compute Padé approximants of $U^2$, for which we can find extrema, without having to worry about convergence issues. For example, to order $(N,M)=(2,4)$, we have
\begin{align}
U^2_{2,4}=\frac{\rho^2\uva{2}^3}{\uva{2}^2+\rho^4\uva{4}^2-\uva{2}\rho^2(\uva{4}+\rho^2\uva{6})}\,,
\end{align}
which has a maximum at 
\begin{align}
\rho_\text{ps}=\frac{\sqrt{\uva{2}}}{(\uva{4}^2-\uva{2}\uva{6})^{1/4}}\,.
\end{align}
In order to obtain a better intuition about this solution, we consider again a scaling of all coefficients of the form (\ref{IllumScaling}), in which case
\begin{align}
\frac{\rho_{\text{ps}}}{2\ma}=\left({\frac{15}{43}}\right)^{1/4}\left[2+43\mathfrak{c}\left(86-\frac{88}{3}\,\frac{\mathfrak{t}_2}{\mathfrak{c}}-440\,\frac{\mathfrak{t}_4}{\mathfrak{c}}-480\,\frac{\mathfrak{t}_6}{\mathfrak{c}}+\frac{864}{5}\,\frac{\mathfrak{x}_2}{\mathfrak{c}}-8\,\frac{\mathfrak{x}_4}{\mathfrak{c}}-96\,\frac{\mathfrak{x}_6}{\mathfrak{c}}\right)+\mathcal{O}(\mathfrak{c}^2)\right]\,,\label{PhotonRadiusPade}
\end{align}
Notice, since the classical position of the photon radius is
\begin{align}
&\rps^{\text{class}}=3\,\ma\,,&&\text{or equivalently} &&\rho_{\text{ps}}^{\text{class}}=(\sqrt{3}+2\text{arccoth}\sqrt{3})\,\ma\sim 3.04901\,\ma\,,\label{ClassSchwarzschildRad}
\end{align}
while $4 \left({\frac{15}{43}}\right)^{1/4}\sim 3.07408$, the result (\ref{PhotonRadiusPade}) is not exact, even in the undeformed case (\emph{i.e.} for the classical Schwarzschild black hole). This is again due to corrections of the type \emph{(i)} discussed in the previous Subsection: they are due to the fact that we are using a Padé approximation in the first place and can be made small by increasing $(N,M)$. This, however, comes at an increased computational complexity, notably since in the terms of order $\mathcal{O}(\mathfrak{c})$, more and more of the coefficients $\{\mathfrak{t}_{2n}\}$ and $\{\mathfrak{x}_{2n}\}$ need to be taken into account.

Fortunately, in the case of the photon radius, there is a different approximation which avoids corrections of order \emph{(i)} and  only captures corrections of type \emph{(ii)} that are due to the deformations of the black hole geometry: indeed, as a function of $\co$, the undeformed potential of the photon radius (\emph{i.e.} for the case $\Psi=\Phi=2\ma$)
\begin{align}
U^2_{\text{class}}=\frac{1-\frac{2\ma}{\co}}{\co^2}=\frac{\co-2\ma}{\co^3}\,,
\end{align}
is in fact a rational function in $\co-\zh=\co-2\ma$. Therefore, any Padé approximation with $N\geq 1$ and $M\geq 3$ is in fact exact for the classical Schwarzschild black hole. For quantum deformed black holes, corrections of the type \emph{(i)} are therefore also absent beyond this order of approximation and we are only left with corrections of the type \emph{(ii)} that depend on (finitely many of) the physical deformation coefficients $\{\theta_{2n}\}_{n\in\mathbb{N}^*}$ and $\{\xi_{2n}\}_{n\in\mathbb{N}^*}$. We are thus lead to consider instead of (\ref{PotentialDistanceExpansion}) the following expansion
\begin{align}
&U^2=\sum_{n=1}^\infty v_n\,(\co-\zh)^n\,,&&\text{with} &&\begin{array}{l} v_1=\frac{\hn{1}}{\zh^2}\,, \\[8pt] v_2=\frac{\hn{2}\zh-4\hn{1}}{2\zh^3}\,, \\[8pt] v_3=\frac{18\hn{1}-6\zh \hn{2}+\zh^2 \hn{3}}{6\zh^4}\,,\end{array}\label{PadePotential}
\end{align}
where $\hn{n}$ are the expansion coefficients of the metric deformations in (\ref{fhnExpansion}) that implicitly depend on the physical coefficients $\{\theta_{2k}\}_{k\leq n}$ and $\{\xi_{2k}\}_{k\leq n}$. From this, we can compute a Padé approximation of the (square of the) potential $(U^2)_{N,M}$ in the variable $\co-\zh$ along with the position of the photon radius. Using the scaling (\ref{IllumScaling}), we can write the latter in the form
\begin{align}
\frac{\rps}{\ma}=\sigma_0+\sigma_1\,\mathfrak{c}+\mathcal{O}(\mathfrak{c}^2)\,.
\end{align}
where $\sigma_{0,1}$ depend on the order of approximation $(N,M)$, as can be seen in Table~\ref{Tab:PositionPade}: for given (finite) order $(N,M)$, $\sigma_1$ only depends on $\mathfrak{c}$ and $\mathfrak{t}_{2L}$ up to $2L=2(N+M)$, thus still respecting the same order of physically relevant coefficients.

\begin{table}
\begin{center}
\begin{tabular}{|c|c||c|c|c|}\hline
&&&\\[-10pt]
$N$ & $M$ & $\sigma_0$ & $\sigma_1$ \\[2pt]\hline
&&&\\[-10pt]
2 & 1 & $(1+\sqrt{3})$ & $(1+\sqrt{3})+\frac{4(8-15 \sqrt{3})}{135}\,\frac{\mathfrak{t}_2}{\mathfrak{c}}+\left(\frac{8}{9}-\frac{16}{\sqrt{3}}\right)\,\frac{\mathfrak{t}_4}{\mathfrak{c}}+-\frac{32}{3}\,\frac{\mathfrak{t}_6}{\mathfrak{c}}$\\[4pt]\hline
%%%
&&&\\[-8pt]
2 & 2 & $\frac{23+9\sqrt{3}}{13}$ & $\frac{23+9\sqrt{3}}{13}-\frac{44(28+33\sqrt{3})}{4095}\frac{\mathfrak{t}_2}{\mathfrak{c}}-\frac{8(3359+1272\sqrt{3})}{2535}\frac{\mathfrak{t}_4}{\mathfrak{c}}-\frac{32(281-24\sqrt{3})}{169}\frac{\mathfrak{t}_6}{\mathfrak{c}}-\frac{768(14-3\sqrt{3})}{169}\frac{\mathfrak{t}_8}{\mathfrak{c}}$\\[6pt]\hline
   %%%
&&&\\[-8pt]
2 & 3 & $3$ & $3-\frac{13471}{14175}\frac{\mathfrak{t}_2}{\mathfrak{c}}-\frac{56128 }{2835}\frac{\mathfrak{t}_4}{\mathfrak{c}}-\frac{3104 }{45}\frac{\mathfrak{t}_6}{\mathfrak{c}}-\frac{1088}{9}\frac{\mathfrak{t}_8}{\mathfrak{c}}-\frac{256 }{3}\frac{\mathfrak{t}_{10}}{\mathfrak{c}}$\\[4pt]\hline
   %%%
&&&\\[-8pt]
3 & 3 & $3$ & $3-\frac{443939}{467775}\frac{\mathfrak{t}_2}{\mathfrak{c}}-\frac{2625104}{127575}\frac{\mathfrak{t}_4}{\mathfrak{c}}-\frac{732448}{8505}\frac{\mathfrak{t}_6}{\mathfrak{c}}-\frac{92608}{405}\frac{\mathfrak{t}_8}{\mathfrak{c}}-\frac{9472}{27}\frac{\mathfrak{t}_{10}}{\mathfrak{c}}-\frac{2048 }{9}\frac{\mathfrak{t}_{12}}{\mathfrak{c}}$\\[4pt]\hline
\end{tabular}
\end{center}
\caption{Approximations of the position of the maximum of $(U^2)_{N,M}$ for different values of $(N,M)$, using the scaling~(\ref{IllumScaling}).}
\label{Tab:PositionPade}
\end{table}

With the position $\rps$, we can next calculate the radius of the black hole shadow in (\ref{bphdef}), which, assuming again the scaling (\ref{IllumScaling}), we expand as
\begin{align}
\frac{\bsh}{\ma}= \tau_0+\tau_1\,\mathfrak{c}+\mathcal{O}(\mathfrak{c}^2)\,.\label{ShadowExpans}
\end{align}
The expressions for $\tau_{0,1}$ depend on the order of the Padé approximation, as is shown in Table~\ref{Tab:ValuePade}. The corrections $\tau_1$ respect the same dependence on the number of physically relevant metric deformation coefficients as $\sigma_1$.

\begin{table}
\begin{center}
\begin{tabular}{|c|c||c|c|c|}\hline
&&&\\[-10pt]
$N$ & $M$ & $\tau_0$ & $\tau_1$ \\[2pt]\hline
&&&\\[-8pt]
2 & 1 & $2(1+\sqrt{3})$ & $2(1+\sqrt{3})+\frac{8(82+75\sqrt{3})}{135}\,\frac{\mathfrak{t}_2}{\mathfrak{c}}+\frac{16(8+3\sqrt{3})}{9}\,\frac{\mathfrak{t}_4}{\mathfrak{c}}+\frac{64}{3}\,\frac{\mathfrak{t}_6}{\mathfrak{c}}$\\[4pt]\hline
%%%
&&&\\[-8pt]
2 & 2 & $2\sqrt{\frac{10}{3}+2\sqrt{3}}$ & \parbox{15cm}{$2\sqrt{\frac{10}{3}+2\sqrt{3}}+\frac{4}{945}\sqrt{\frac{12051454}{3}+2415638\sqrt{3}}\frac{\mathfrak{t_2}}{\mathfrak{c}}+\frac{88}{45}\sqrt{\frac{265}{3}+59\sqrt{3}}\frac{\mathfrak{t}_4}{\mathfrak{c}}+\frac{32}{3}\sqrt{38\sqrt{3}-\frac{146}{3}}\frac{\mathfrak{t}_6}{\mathfrak{c}}+ 128\sqrt{\sqrt{3}-\frac{5}{3}}\frac{\mathfrak{t}_8}{\mathfrak{c}}$}\\[6pt]\hline
   %%%
&&&\\[-8pt]
2 & 3 & $3\sqrt{3}$ & $3\sqrt{3}+\frac{1}{\sqrt{3}}\left(\frac{98837}{4725 }\frac{\mathfrak{t}_2}{\mathfrak{c}}+\frac{136096}{2835 }\frac{\mathfrak{t}_4}{\mathfrak{c}}+\frac{4384}{45 }\frac{\mathfrak{t}_6}{\mathfrak{c}}+\frac{1216}{9} \frac{\mathfrak{t}_8}{\mathfrak{c}}+\frac{256 }{3}\frac{\mathfrak{t}_{10}}{\mathfrak{c}2}\right)$\\[4pt]\hline
   %%%
&&&\\[-8pt]
3 & 3 & $3\sqrt{3}$ & $3\sqrt{3}+\frac{1}{\sqrt{3}}\left(\frac{652294}{31185}\frac{\mathfrak{t}_2}{\mathfrak{c}}+\frac{2066276}{42525}\frac{\mathfrak{t}_4}{\mathfrak{c}}+\frac{62528}{567}\frac{\mathfrak{t}_6}{\mathfrak{c}}+\frac{29152}{135}\frac{\mathfrak{t}_8}{\mathfrak{c}}+\frac{2560}{9}\frac{\mathfrak{t}_{10}}{\mathfrak{c}}+\frac{512}{3}\frac{\mathfrak{t}_{12}}{\mathfrak{c}}\right)$\\[4pt]\hline
\end{tabular}
\end{center}
\caption{Approximations of the value of the black hole shadow stemming from and approximation of the potential $U^2\sim (U^2)_{N,M}$ for different values of $(N,M)$, using the scaling (\ref{IllumScaling}).}
\label{Tab:ValuePade}
\end{table}

For approximations of the potential of the form $(U^2)_{N,3}$ we can in fact write the (approximate) result in an even more compact form. To this end, we write $\tau_1$ as
\begin{align}
\tau_1(N)=3\sqrt{3}+\sum_{k=1}^{3+N}\beta_{2k}(N)\,\frac{\mathfrak{t}_{2k}}{\mathfrak{c}}\,,\label{TauParametrisation}
\end{align}
where $\beta_{2k}$ are numerical coefficients (which for $N=2,3$ can be found in Table~\ref{Tab:ValuePade}). For $N$ sufficiently large, a fixed $\beta_{2k}$ numerically changes very little when $N$ is increased: 
\begin{figure}
    \centering
\includegraphics[width=0.5\textwidth]{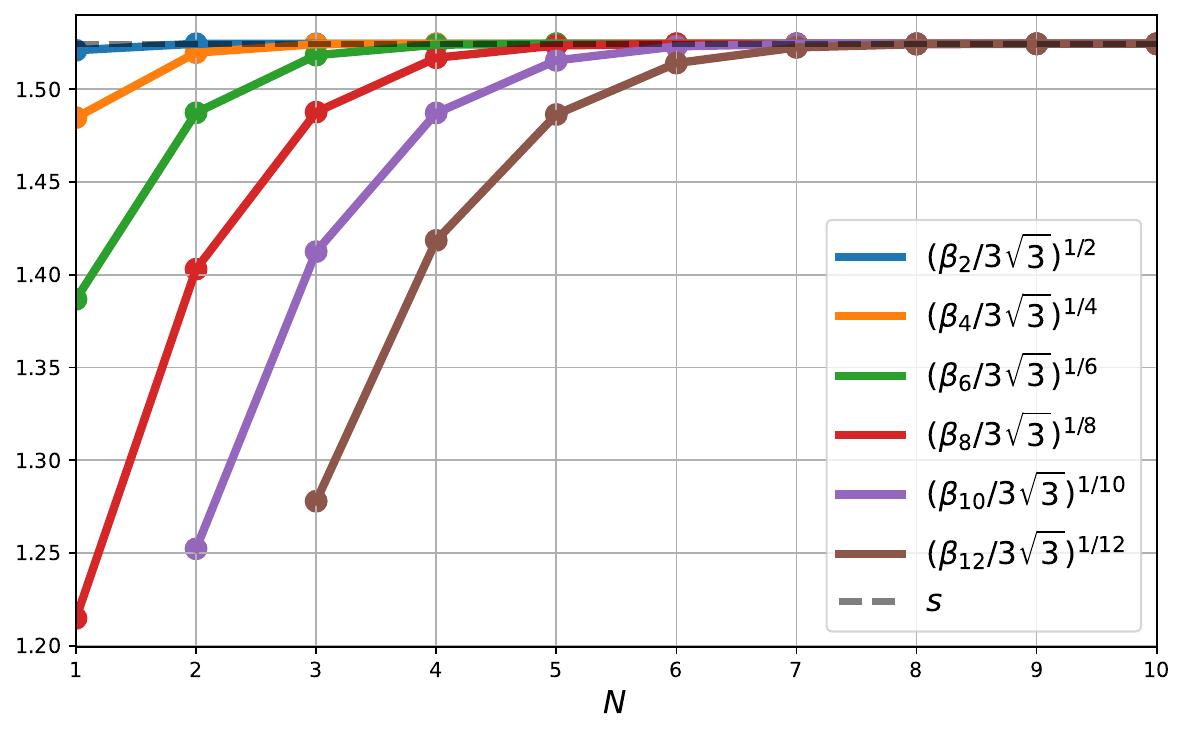}
    \caption{Coefficients $\beta_{2n}$ in the parametrisation of $\tau_1$ in (\ref{TauParametrisation}) for different Padé approximations $(N,3)$.}
    \label{fig:ConvergenceCoefs}
\end{figure}
as is shown numerically in Figure~\ref{fig:ConvergenceCoefs},  $\beta_{2k}(N)$ approaches asymptotically a fixed numerical value, namely
\begin{align}
&\lim_{N\to \infty}\beta_{2k}(N)\to 3\sqrt{3}\,\mathfrak{s}^{2k}\hspace{0.5cm}\forall k\geq 1\,,&&\text{with} &&\mathfrak{s}\sim 1.5245\sim \frac{\rho^{\text{class}}_{\text{ps}}}{2\ma}\,.\label{Svalue}
\end{align}
We notice that $2\ma\mathfrak{s}$ is indeed numerically very close to the proper distance of the photon sphere in the case of the (classical)  Schwarzschild black hole (see (\ref{ClassSchwarzschildRad})). Assuming that the observation (\ref{Svalue}) indeed holds $\forall k\geq 1$, we can write the asymptotic behaviour of $\tau_1(N)$ in the following form
\begin{align}
\lim_{N\to \infty}(\mathfrak{c}\,\tau_1(N))=3\sqrt{3}\,\mathfrak{c}+3\sqrt{3}\,\sum_{k=1}^\infty\,\mathfrak{s}^{2k}\,\mathfrak{t}_{2k}=3\sqrt{3}\,\mathfrak{c}+3\sqrt{3}\,\sum_{k=1}^\infty\,(2\ma\mathfrak{s})^{2k}\,\theta_{2k}\,,
\end{align}
where we have used (\ref{IllumScaling}). Using furthermore the series expansion (\ref{SeriesExpansionMetric}) we can formally write
\begin{align}
\lim_{N\to \infty}(\mathfrak{c}\,\tau_1(N))=3\sqrt{3}\,\mathfrak{c}+\frac{3\sqrt{3}}{2\ma}\,\left(\Psi(2\ma\,\mathfrak{s})-\zh\right)\,,
\end{align}
where we have implicitly assumed that $2\ma\,\mathfrak{s}$ is inside the radius of convergence of (\ref{SeriesExpansionMetric}).\footnote{A different way of phrasing this assumption is that $\displaystyle \lim_{N\to \infty} \Psi_{N,3}(2\ma\mathfrak{s})=\Psi(2\ma\mathfrak{s})-\zh$. We note that we have analysed the soundness of this assumption in a number of models of quantum black holes in Appendix~\ref{Sect:Convergence} and find it to be justified at least for small values of the parameter that governs the size of quantum corrections.} Thus, using (\ref{RescaledMetricFunctions}) and (\ref{ShadowExpans}), to leading order, the approximate black hole shadow can be written entirely in terms of the deformation function $\Psi$ evaluated at a particular value
\begin{align}
\lim_{N\to \infty}\frac{\bsh(N)}{\ma}=\frac{\bsh}{\ma}\sim\frac{3\sqrt{3}}{2\ma}\Psi(2\ma\,\mathfrak{s})\,.\label{CompactCorrectionShadow}
\end{align}
Turning this result around, we can use (\ref{CompactCorrectionShadow}) to provide a bound for the size of the deformation function based on an experimental observation of the photon radius. Indeed, assuming absence of rotation (which is compatible with the experimental data), at a 1-$\sigma$ confidence level, the EHT observations of M87$^\ast$ lead to \cite{EventHorizonTelescope:2021dqv}
\begin{equation}
    \tilde{b}_\text{sh}^{\text{EHT M}87^\ast \, (\pm)} = 3 \sqrt{3}M_{\text{M}87^\ast}(1 \pm 0.17) \ ,
\end{equation}
where $M_{\text{M}87^\ast}$ is the mass of the black hole M87$^\ast$. Thus, assuming M87$^\ast$ to be spherically symmetric, we find based on \eqref{CompactCorrectionShadow}, the following rough bound for the function governing its deviation from a Schwarzschild black hole
\begin{equation}
 \abs{\frac{\Psi(2\ma_{\text{M}87^\ast}\,\mathfrak{s})}{2M_{\text{M}87^\ast}} - 1} \leq 0.17 \ .\label{ExperimentalBound}
\end{equation}

%%%%%%%%%%%%%%%%%%%%%%%%%%%%%%%%%%%%%%%%%%%%%%%

\section{Simple Examples of Corrected Schwarzschild Geometries}\label{subs example models}
In order to illustrate and showcase the formalism developed in the previous Section, we shall consider a number of models of spherically symmetric and static black holes that have previously been proposed in the literature \cite{Hayward_2006,Simpson:2018tsi,Simpson:2019mud,Dymnikova:1992ux,Garcia95,dilaton}. These examples have been selected on the basis that they are compatible with (\ref{IllumScaling}), such that we can use the effective expansions to leading order in the quantum deformations. Furthermore, these black holes display a wide variety on different functional dependencies on the Schwarzschild coordinate $r$ and an explicit parameter (called $\epsilon$) that governs the size of quantum effects. In all cases, following the philosophy of our effective approach, the theory of quantum gravity in which context these models have been proposed, plays no role.

\subsection{Metrics and Deformation Parameters}

We start with four geometries \cite{Hayward_2006,Simpson:2018tsi,Simpson:2019mud,Dymnikova:1992ux} that have been proposed to describe spherically symmetric and static black holes, as deformations of the Schwarzschild geometry \cite{Wald:1984rg}. We label them as 'simple' in the sense that they are characterised by $f(\co)=h(\co)$ $\forall \co\geq \zh$ in (\ref{MetricGeneral}).

In order to describe the four models proposed in \cite{Hayward_2006,Simpson:2018tsi,Simpson:2019mud,Dymnikova:1992ux} in parallel (as far as this is possible), we first introduce a unifying notation: in addition to the mass of the black hole $\ma$, all four models introduce an additional parameter, which we shall call $\eta$, to describe the deformation of the Schwarzschild space-time. While $\eta$ is conceptually quite different in all models, we shall in the following write it in terms of a different small parameter, which can be uniformly defined. To introduce the latter, we first define 
\begin{equation}
x:=\frac{\co}{2\ma} \qq{and} \xh:=\frac{\zh}{2\ma}\,.
\end{equation}
in order to absorb the mass-scale in the radial coordinate. We then define the fractional deviation of the event horizon radius from the Schwarzschild radius,\footnote{For $\epsilon\ll 1$, this parametrisation ensures that we are not expanding the metric around the internal horizon that arises with the regularisation of the singularity at the origin within the models examined, but rather around the external horizon, which is expected to be close the classical Schwarzschild radius $\xh^{\text{class}}=1$.} as in \cite{Rezzolla:2014mua} 
\begin{equation}
    \epsilon:= \frac{2M-\hr}{\hr} = \frac{1-\xh}{\xh} \ \longleftrightarrow \ \xh = \frac{1}{1+\epsilon}\ .\label{EpsEq}
\end{equation}
With this notation set up, Table~\ref{subs models table} presents the metric deformation functions for the four models, along with the parameter $\eta$ as a function of $\epsilon$, which can take the form of a rational, exponential, or non-analytic function at $\eta = 0$. In anticipation of our further discussion, the table also indicates the leading correction of the position of the photon ring and the black hole shadow as functions of $\epsilon$.

%Here we show, as a benchmark of the Padé approximation, the computation of the photon sphere radius in three different black hole models. Each metric is characterised by a deformation function that depends on a free parameter $\eta$ (rational, exponential and non-analytic function at $\eta=0$). We report the main features of the models in Table \ref{subs models table}.
\begin{table}[h!]
\centering
\begin{tabular}{c|c|c|c|c} 
\toprule
Model          & $\displaystyle f=h$                             & $\displaystyle \eta(\epsilon)$                                                                  & $ \rps/M-3$  & $\bsh /M - 3\sqrt{3}$\\[3pt]
\midrule
Hayward \cite{Hayward_2006}    & $\displaystyle 1-\frac{x^2}{x^3+\eta}$        & $\displaystyle\frac{\epsilon}{(1+\epsilon)^3}$                                                   & $-\dfrac{16}{9}\epsilon + \order{\epsilon^2}$        & $-\dfrac{8}{3\sqrt{3}}\epsilon + \order{\epsilon^2} $         \\[8pt]
Simpson-Visser (I) \cite{Simpson:2018tsi} & $\displaystyle1-\frac{1}{\sqrt{x^2+\eta}}$    & $\displaystyle\frac{\epsilon\,(2+\epsilon)}{(1+\epsilon)^2}$                                             & $-\dfrac{20}{9}\epsilon + \order{\epsilon^2}$       & $-\dfrac{4}{\sqrt{3}}\epsilon + \order{\epsilon^2}$             \\[11pt]
Simpson-Visser (II) \cite{Simpson:2019mud} & $\displaystyle1-\frac{1}{x}e^{-\eta / x}$    & $\displaystyle\frac{\log(1+\epsilon)}{1+\epsilon}$                                             & $-\dfrac{8}{3}\epsilon + \order{\epsilon^2}$            & $-2\sqrt{3} \, \epsilon + \order{\epsilon^2}$        \\[8pt]
Dymnikova \cite{Dymnikova:1992ux}     & $\displaystyle1-\frac{1-e^{-x^3 / \eta}}{x}$ & $\displaystyle-\left[(1+\epsilon)^3 \log \left(\frac{\epsilon}{1+ \epsilon} \right)\right]^{-1}$ & $ \left(\dfrac{81}{8}\log \epsilon - 3 \right) \epsilon^{27/8} + \order{\epsilon^{33/8}}$ & $- 3  \sqrt{3} \, \epsilon^{27/8} + \order{\epsilon^{33/8}}$\\[8pt]
\bottomrule
\end{tabular}
\caption{Summary of the geometries of the four models studied in this Section.}
\label{subs models table}
\end{table}
Notice that in all models of this type, the deformation coefficients $\{\mathfrak{x}_{2n}\}_{n\in\mathbb{N}}$ in (\ref{RescaledMetricFunctions}) (which are identical to $\{\mathfrak{t}_{2n}\}_{n\in\mathbb{N}}$), do not explicitly depend on the mass: in fact, from (\ref{DiffEqDistance}) it follows that $\rho(r)=2\,\ma\,\underline{\rho}(x,\epsilon)$, where the function $\underline{\rho}(x,\epsilon)$ does not explicitly depend on $\ma$. From (\ref{hfgen}) and (\ref{SeriesExpansionMetric}), we can then write
\begin{align}
f=h=1-\frac{\xh+\sum_{n=1}^\infty \xi_{2n}\,(2\ma)^{2n}\,\underline{\rho}(x,\epsilon)^{2n}}{x}\ ,
\end{align}
such that, in order for $f$ to be independent of $\ma$, we require $\xi_{2n}=\frac{\mathfrak{x}_{2n}(\epsilon)}{(2M)^{2n}}$, as in (\ref{RescaledMetricFunctions}). Furthermore, from (\ref{EpsEq}), we have $\mathfrak{c}=-\frac{\epsilon}{1+\epsilon}$ and we shall see that in all cases $\mathfrak{x}_{2n}=\mathcal{O}(\epsilon)$ (such that $\frac{\mathfrak{x}_{2n}}{\mathfrak{c}}=\mathcal{O}(1)$), compatible, with (\ref{IllumScaling}). In the following we shall discuss the four models in more details.\\

%%%% COMPARISON x_n HAYWARD SV1 SV2
\begin{figure}[t]
\begin{center}
\includegraphics[width=\textwidth]{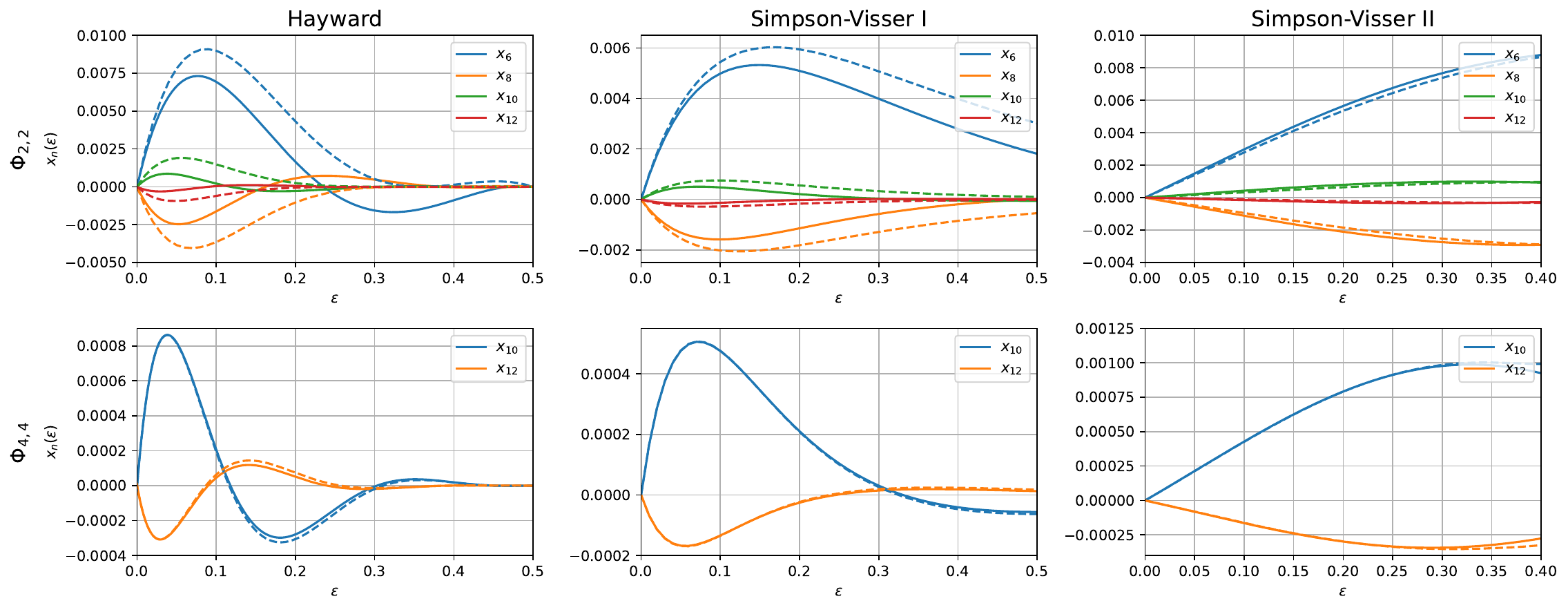}
\end{center}
\caption{Comparison of the deformation coefficients of the metric $\mathfrak{x}_{2n}$ (solid lines) with the (mass-rescaled) coefficients (\ref{ReplacementCoefficients}) stemming from a Padé approximation of $\Phi$ (dashed lines) as functions of $\epsilon$. Top row: the coefficients $\mathfrak{x}_{6,8,10,12}$ and their approximations stemming from the Padé approximant $\Phi_{2,2}$ of for the Hayward metric (left column), Simpson-Visser I metric (middle column) and Simpson-Visser II metric (right column). Bottom row: the coefficients $\mathfrak{x}_{10,12}$ and their approximations stemming from the Padé approximant $\Phi_{4,4}$.}
\label{Fig:XiReplacementHayward}
\end{figure}

%%%% COMPARISON U^2 HAYWARD SV1 SV2
\begin{figure}[!h]
\centering
\includegraphics[width=\textwidth]{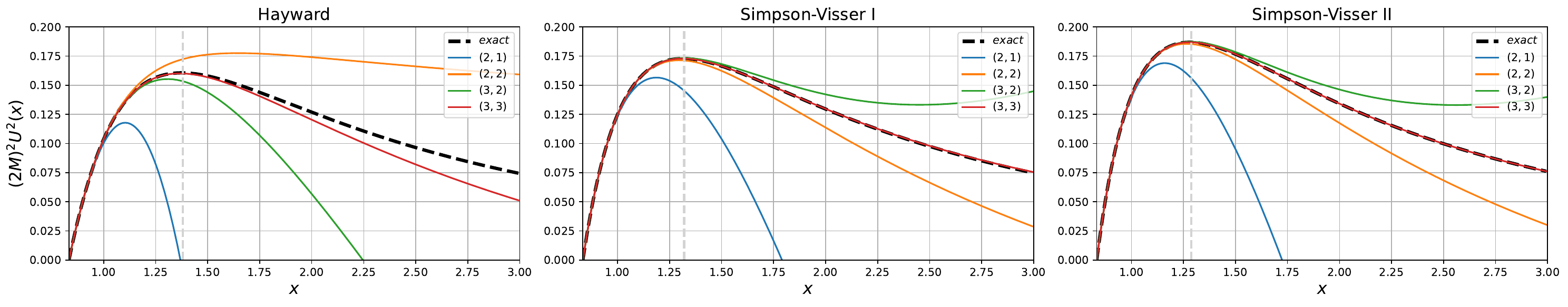}
\caption{Comparison of the potential $U^2$ for Hayward (left panel), Simpson-Visser I (middle panel), and Simpson-Visser II (right panel) with various Padé approximants \eqref{PadePotential} for each model with a fixed $\epsilon=0.2$. The vertical dashed gray line denotes the position of the maximum of the (exact) potential. } 
\label{Fig:PotentialPlots}
\end{figure}

\begin{figure}[t]
\centering
\includegraphics[width=\textwidth]{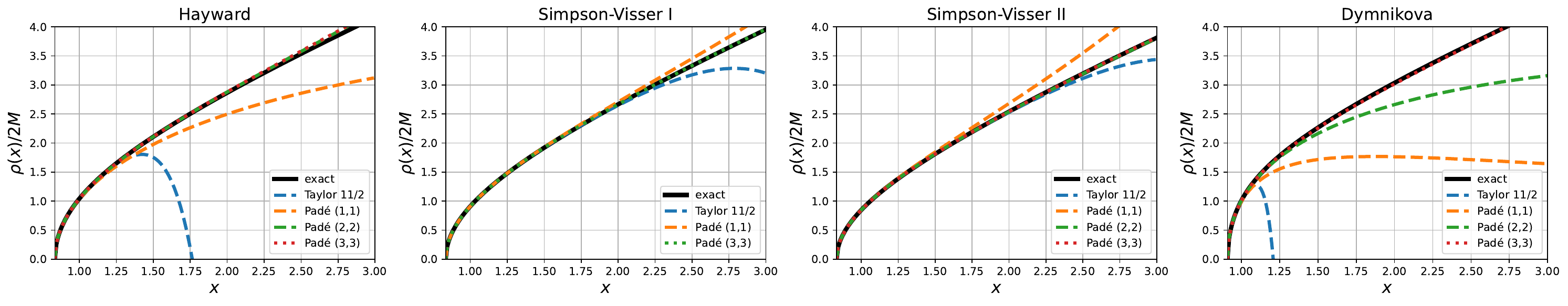}
\caption{Comparison of approximations of the distance (as a function of $x$) in the case of the Hayward, Simpson-Visser I and Simpson-Visser II geometries (for $\epsilon=0.2$) and Dymnikova (for $\epsilon=0.1$): the solid black lines represent the numerical integration of (\ref{DiffEqDistance}), while the dashed blue lines represent the Taylor series expansion (\ref{RhoSolution}) (including terms up to order $n=6$) and the coloured dashed and dotted lines show Padé approximants of the distance as in (\ref{PadeDistance}).} 
\label{Fig:DistanceHayward}
\end{figure}

\textbf{Hayward black hole}: we start with the Hayward black hole, which was first introduced in \cite{Hayward_2006}. Rescaling the deformation functions of the metric as in (\ref{RescaledMetricFunctions}), we find explicitly for the leading ones (with $\mathfrak{t}_{2n}=\mathfrak{x}_{2n}$ $\forall n\geq 1$)
\begin{align}
&\mathfrak{x}_2=\frac{3 (1-2 \epsilon ) \epsilon }{4 (1+\epsilon )}\,,&&\mathfrak{x}_{4}=-\frac{\epsilon  (1-2 \epsilon ) (\epsilon  (7 \epsilon -22)+7)}{16 (1+\epsilon )}\,,&&\mathfrak{x}_6=\frac{\epsilon  (1-2 \epsilon ) (\epsilon  (\epsilon  (2 \epsilon  (61 \epsilon -899)+3225)-1600)+221)}{960 (1+\epsilon )}\ .
\end{align}
In a first step, we can compare these coefficients $\{\mathfrak{x}_{2n}\}$ with their counterparts stemming from a Padé approximant $\Phi_{N,M}$($=\Psi_{N,M}$) as in (\ref{PadeApproximation}): Figure~\ref{Fig:XiReplacementHayward} (left column) shows a comparison of the exact coefficients and their approximations for $(N,M)=(2,2)$ (upper panel) and $(N,M)=(4,4)$ (lower panel). While in the first case, the approximations capture at least roughly the functional dependence with respect to $\epsilon$, in the latter case, the $\widetilde{\mathfrak{x}}_{2n}$ even present good approximations for finite values of $\epsilon$.\\
In Figure~\ref{Fig:PotentialPlots} (left panel) we show the comparison with the analytical form of the effective potential $U^2$ with the Padé approximations of various orders.
Similarly, Figure~\ref{Fig:DistanceHayward} (left) compares different approximations of the distance function (for a fixed value of $\epsilon=0.2$) as functions of $x$: the plot highlights the finite radius of convergence of the simple series expansion (\ref{RhoSolution}) and also shows that viable approximations can already be obtained at fairly low order of the Padé approximant. The position of the maximum of the potential as well as the shadow following tables~\ref{Tab:PositionPade} and \ref{Tab:ValuePade} respectively, is given as follows
\begin{center}
\begin{tabular}{|c|c||c|c|c|}\hline
&&&\\[-10pt]
$N$ & $M$ & $\sigma_1\,\mathfrak{c}$ & $\tau_1\,\mathfrak{c}$ \\[2pt]\hline
&&&\\[-8pt]
2 & 1 & $-\frac{11 - 3 \sqrt{3}}{3}\,\epsilon+\mathcal{O}(\epsilon^2)$ &  $\frac{1-3\sqrt{3}}{3}\,\epsilon+\mathcal{O}(\epsilon^2)$\\[4pt]\hline
%%%
&&&\\[-10pt]
2 & 2 & $\frac{3}{169}(-131+16\sqrt{3})\,\epsilon+\mathcal{O}(\epsilon^2)$ & $-\frac{1}{6}\sqrt{-\frac{965}{3}+233\sqrt{3}}\,\epsilon +\mathcal{O}(\epsilon^2)$\\[4pt]\hline
   %%%
&&&\\[-8pt]
2 & 3 & $-\frac{16}{9}\,\epsilon+\mathcal{O}(\epsilon^2)$ & $-\frac{8}{3\sqrt{3}}\,\epsilon+\mathcal{O}(\epsilon^2)$ \\[4pt]\hline
   %%%
&&&\\[-8pt]
3 & 3 & $-\frac{16}{9}\,\epsilon+\mathcal{O}(\epsilon^2)$ & $-\frac{8}{3\sqrt{3}}\,\epsilon+\mathcal{O}(\epsilon^2)$\\[4pt]\hline
\end{tabular}
\end{center}

\textbf{Simpson-Visser I black hole:} as a second example, we examine the geometry introduced in \cite{Simpson:2019mud} and also in \cite{Cunha:2020azh,Culetu:2014lca}. The leading deformation coefficients are given by
\begin{equation}
    \mathfrak{x}_2 = \frac{\epsilon (2 + \epsilon)}{4 (1+\epsilon)^3} \  , \quad \mathfrak{x}_4 = \frac{\epsilon(2 + \epsilon)(\epsilon (2 + \epsilon)-11)}{96(1 + \epsilon)^5} \ , \quad
    \mathfrak{x}_6 = \frac{\epsilon  (\epsilon +2) (\epsilon  (\epsilon +2) (\epsilon  (\epsilon +2)-121)+292)}{5760 (\epsilon +1)^7} \ .
\end{equation}
Conclusions similar to Hayward's can be drawn from Figure~\ref{Fig:XiReplacementHayward} (middle column), where the comparison of the $\{\mathfrak{x}_{6}, \mathfrak{x}_{8}, \mathfrak{x}_{10}, \mathfrak{x}_{12}\}$ with their approximations computed with a Padé approximant with $(N,M)=(2,2)$ (upper panel) and $\{ \mathfrak{x}_{10}, \mathfrak{x}_{12}\}$ with their approximation using a Padé approximation with $(N,M)=(4,4)$ (lower panel). In Fig.~\ref{Fig:DistanceHayward} (second panel from the left) we compare the proper distance numerically computed with the exact metric and the Padé approximants of different orders.
In Figure~\ref{Fig:PotentialPlots} (middle panel) we show the comparison with the analytical form of the effective potential $U^2$ with the Padé approximations of various orders.
Finally, in Figure~\ref{Fig:DistanceHayward} (middle) we compare different approximations of the distance function numerically integrated (for a fixed value of $\epsilon=0.2$) and the Padé approximants.\footnote{We noticed that Padé of order $(M,N)=(2,2)$ introduces a vertical tangent behaviour at a position $\tilde{x} \geq \xh$ which is a feature purely due to the approximation and is effectively eliminated with a suitable choice of the Padé approximant order.}

The position of the maximum of the potential as well as the shadow following tables~\ref{Tab:PositionPade} and \ref{Tab:ValuePade} respectively, is given as follows
\begin{center}
\begin{tabular}{|c|c||c|c|c|}\hline
&&&\\[-10pt]
$N$ & $M$ & $\sigma_1\,\mathfrak{c}$ & $\tau_1\,\mathfrak{c}$ \\[2pt]\hline
&&&\\[-8pt]
2 & 1 & $-\frac{13}{6}\,\epsilon+\mathcal{O}(\epsilon^2)$ &  $-\left(\frac{2}{3}+\sqrt{3}\right)\,\epsilon+\mathcal{O}(\epsilon^2)$\\[4pt]\hline
%%%
&&&\\[-10pt]
2 & 2 & $-\frac{15}{338}(47+2\sqrt{3})\,\epsilon+\mathcal{O}(\epsilon^2)$ & $-\frac{1}{6}\,\sqrt{\frac{265}{3}+59\sqrt{3}}\,\epsilon +\mathcal{O}(\epsilon^2)$\\[4pt]\hline
   %%%
&&&\\[-8pt]
2 & 3 & $-\frac{20}{9}\,\epsilon+\mathcal{O}(\epsilon^2)$ & $-\frac{4}{\sqrt{3}}\,\epsilon+\mathcal{O}(\epsilon^2)$ \\[4pt]\hline
   %%%
&&&\\[-8pt]
3 & 3 & $-\frac{20}{9}\,\epsilon+\mathcal{O}(\epsilon^2)$ & $-\frac{4}{\sqrt{3}}\,\epsilon+\mathcal{O}(\epsilon^2)$\\[4pt]\hline
\end{tabular}
\end{center}
Before moving to examples with exponential dependence on the radial coordinate, we remark that the black hole geometries proposed by Bardeen \cite{bardeen1968}, Frolov \cite{Frolov:PhysRevD.94.104056} and the one proposed in \cite{Cadoni:2022vsn} have a similar functional dependence on $x$ to the ones of Hayward and Simpson-Visser I. We therefore expect to obtain similar results in these cases as reported above.\\

\textbf{Simpson-Visser II black hole:} for the third model, we study the metric introduced in \cite{Simpson:2018tsi}, where the deformation function is an exponential of the inverse the radial coordinate. This metric is constructed in such a way that at the origin a Minkowski core is recovered rather than a de Sitter one. We report the leading deformation coefficients
\begin{gather}
    \mathfrak{x}_2 = -\frac{\epsilon +1}{4}  (\log (\epsilon +1)-1) \log (\epsilon +1) \  , \quad \mathfrak{x}_4 = \frac{(\epsilon +1)^3}{96}  (\log (\epsilon +1)-1) \log (\epsilon +1) (\log (\epsilon +1) (4 \log (\epsilon +1)-13)+8) \ , \\
    \mathfrak{x}_6 = -\frac{(\epsilon +1)^5}{5760} (\log (\epsilon +1)-1) \log (\epsilon +1) (\log (\epsilon +1) (\log (\epsilon +1) (\log (\epsilon +1) (34 \log (\epsilon +1)-263)+626)-568)+172)\ .
\end{gather}
In Figure~\ref{Fig:XiReplacementHayward} (right column) we show the comparison of the coefficients $\{\mathfrak{x}_{6}, \mathfrak{x}_{8}, \mathfrak{x}_{10}, \mathfrak{x}_{12}\}$ with their approximations computed with a Padé approximant with $(N,M)=(2,2)$ (upper panel) and $\{ \mathfrak{x}_{10}, \mathfrak{x}_{12}\}$ with their approximation using a Padé approximation with $(N,M)=(4,4)$ (lower panel). In Fig.~\ref{Fig:DistanceHayward} (third panel from the left) we compare the proper distance numerically computed with the exact metric and the Padé approximants of different orders.
In Figure~\ref{Fig:PotentialPlots} (right panel) we show the comparison with the analytical form of the effective potential $U^2$ with the Padé approximations of various orders.
Finally, in Figure~\ref{Fig:DistanceHayward} (middle) we compare different approximations of the distance function numerically integrated (for a fixed value of $\epsilon=0.2$) and the Padé approximants.\footnote{We notice that the Padé approximant of order $(N,M)=(2,2)$ introduces a vertical tangent behaviour at a point outside of the horizon, which is a feature purely due to the (low order of) approximation and is effectively eliminated with a suitable choice of the Padé approximant order.}

The position of the maximum of the potential as well as the shadow following tables~\ref{Tab:PositionPade} and \ref{Tab:ValuePade} respectively, is given as follows
\begin{center}
\begin{tabular}{|c|c||c|c|c|}\hline
&&&\\[-10pt]
$N$ & $M$ & $\sigma_1\,\mathfrak{c}$ & $\tau_1\,\mathfrak{c}$ \\[2pt]\hline
&&&\\[-8pt]
2 & 1 & $-\frac{2}{3}\,(2+\sqrt{3})\,\epsilon+\mathcal{O}(\epsilon^2)$ &  $-\frac{4}{3}\,(1+\sqrt{3})\,\epsilon+\mathcal{O}(\epsilon^2)$\\[4pt]\hline
%%%
&&&\\[-10pt]
2 & 2 & $-\frac{6}{13}\,(4+\sqrt{3})\,\epsilon+\mathcal{O}(\epsilon^2)$ & $-\frac{4}{3}\,\sqrt{\frac{10}{3}+2\sqrt{3}}\,\epsilon +\mathcal{O}(\epsilon^2)$\\[4pt]\hline
   %%%
&&&\\[-8pt]
2 & 3 & $-\frac{8}{3}\,\epsilon+\mathcal{O}(\epsilon^2)$ & $-2\,\sqrt{3}\,\epsilon+\mathcal{O}(\epsilon^2)$ \\[4pt]\hline
   %%%
&&&\\[-8pt]
3 & 3 & $-\frac{8}{3}\,\epsilon+\mathcal{O}(\epsilon^2)$ & $-2\,\sqrt{3}\,\epsilon+\mathcal{O}(\epsilon^2)$\\[4pt]\hline
\end{tabular}
\end{center}

\textbf{Dymnikova black hole:} special care must be paid to the Dymnikova spacetime \cite{Dymnikova:1992ux}, as its deformation function is non-analytic around $\eta=0$. To address this, we use a double expansion in the power series of $u := e^{-1/\eta}$ and $\log u$. Specifically, the outer event horizon can be expanded as $\xh^{(\text{Dymn})} = 1 - u + 3u^2 \log u + \order{u^3}$. Since therefore the deformations of the position of the photon radius and the black hole shadow are not analytic functions of $\mathfrak{c}$, we cannot directly use the results of the tables~\ref{Tab:PositionPade} and \ref{Tab:ValuePade}. Nevertheless, we can demonstrate the efficiency of using Padé approximants to study the geometry and observables computed from it: we have plotted in the right panel of Figure~\ref{Fig:DistanceHayward} different approximations of the distance function: indeed, at order $(3,3)$, the Padé approximant represents an excellent approximation of the (numerically calculated) distance function (at $\epsilon=0.1$). 
\begin{figure}[t]
    \centering
    \includegraphics[width=\textwidth]{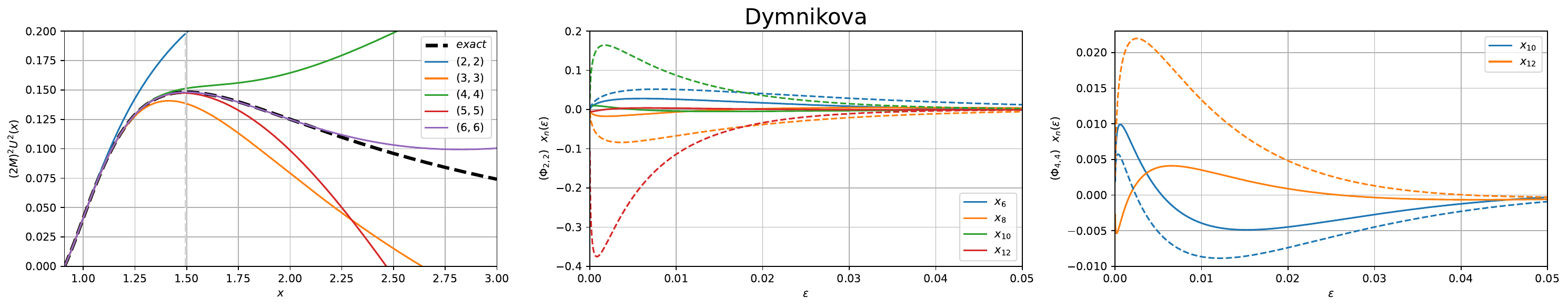}
    \caption{Left panel: comparison of the effective potential $U^2$ for the Dymnikova metric with various Padé approximants \eqref{PadePotential} for fixed $\epsilon=0.1$. The vertical dashed grey line indicates the location of the maximum of the exact potential. Middle and right panels: comparison of the deformation coefficients of the Dymnikova metric $\mathfrak{x}_{2n}$ (solid lines) with the (mass-rescaled) coefficients (\ref{ReplacementCoefficients}) (dashed lines) stemming from a Padé approximation $\Phi_{2,2}$ (middle panel) and $\Phi_{4,4}$ (right panel) as functions of $\epsilon$. }
    \label{fig:dymnpot}
\end{figure}
For completeness, here we report the first two deformation coefficients as in eq.~\eqref{RescaledMetricFunctions} as functions of $\epsilon$, which read
\begin{align}
&\mathfrak{x}_2= -\frac{3 \epsilon}{4}  (\epsilon +1) \log \left(\frac{\epsilon }{\epsilon +1}\right) \left[3 \epsilon  \log \left(\frac{\epsilon }{\epsilon +1}\right)+1\right] \ , \\ &\mathfrak{x}_4 = -\frac{\epsilon}{32}  (\epsilon +1)^3 \log \left(\frac{\epsilon }{\epsilon +1}\right) \left[3 \epsilon  \log \left(\frac{\epsilon }{\epsilon +1}\right)+1\right] \left[9 \log \left(\frac{\epsilon }{\epsilon +1}\right) \left(2 \epsilon +4 \epsilon  \log \left(\frac{\epsilon }{\epsilon +1}\right)+1\right)+4\right] \ ,
\end{align}
and in the middle and right panels in Fig.~\ref{fig:dymnpot} we plot the higher order coefficients with their approximations with Padé of orders respectively $(N,M)=(2,2)$ and $(N,M)=(4,4)$. 
We can see that the expansion around $\epsilon=0$ of such coefficients introduces a behaviour of the form $\mathfrak{x}_{2n} \propto \epsilon \log^{n}(\epsilon)$ which introduces numerical instability. Nevertheless we know that it has an analytical, but yet steep, convergence to 0, as one can see from the plots in Figure~\ref{fig:dymnpot} .

\begin{figure}[t]
    \centering
    \includegraphics[width=0.5\textwidth]{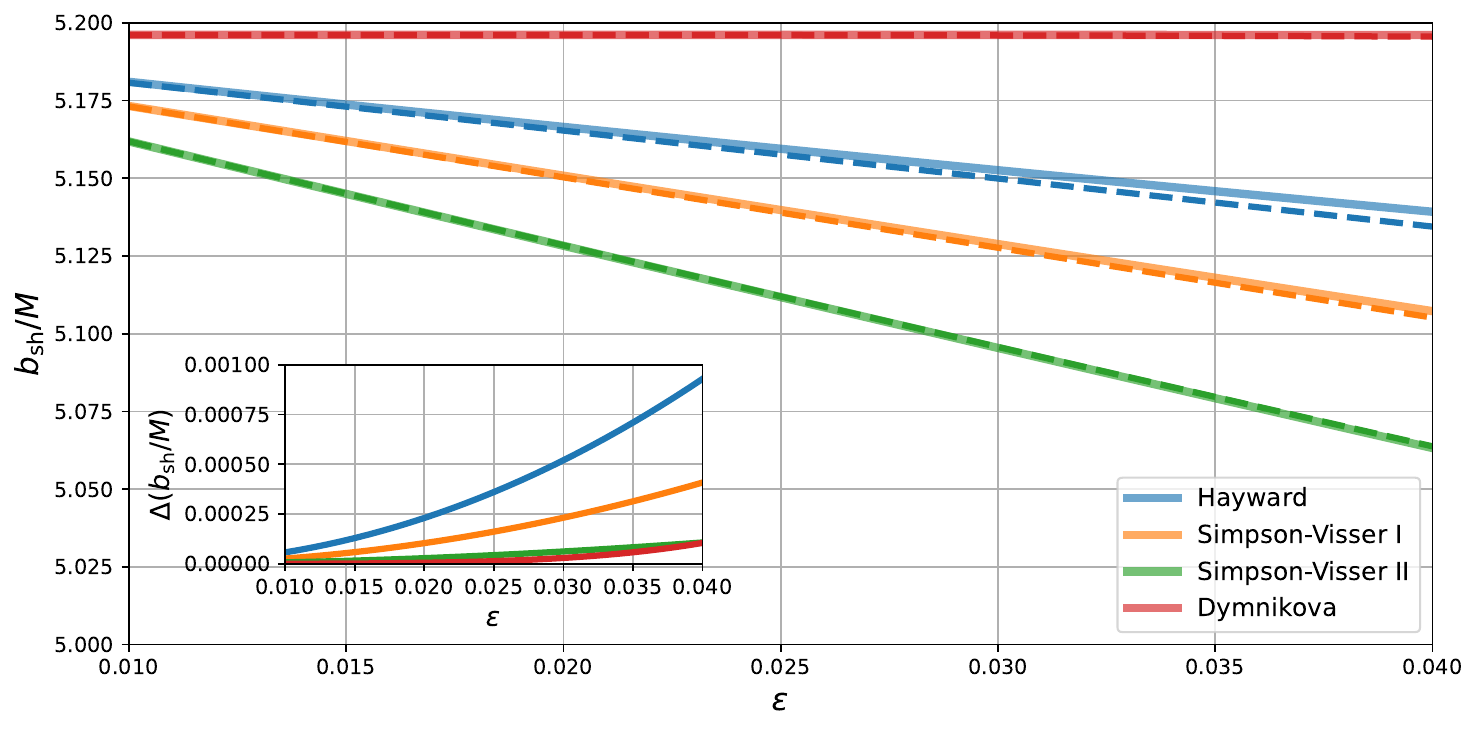}
    \caption{Comparison of the shadow computed from eq.~(\ref{CompactCorrectionShadow}) (dashed lines) to the exact numerical value (solid lines) for the models in Table \ref{subs models table}. The inset plot shows the relative difference as explained in the text.}
    \label{fig: shadow comparison}
\end{figure}

Finally, for all examples we can showcase the effective expression (\ref{CompactCorrectionShadow}) for the black hole shadow, quantum-corrected to leading order: Figure~\ref{fig: shadow comparison} compares the result $\bsh^{\text{app}}/\ma$ obtained from (\ref{CompactCorrectionShadow}) (dashed lines) in each case with the numerical result (solid lines) $\bsh^{\text{exact}}/\ma$, computed from the full metric in each case, as functions of (small) $\epsilon$. In order to appreciate the quality of the approximation furnished by (\ref{CompactCorrectionShadow}), the inset plot shows the relative difference $\Delta (\bsh/\ma)=\left|\frac{\bsh^{\text{exact}}-\bsh^{\text{app}}}{\bsh^{\text{exact}}}\right|$: in all cases, for small values of $\epsilon$, the latter is of the order of $<0.1\%$, indicating that (\ref{CompactCorrectionShadow}) indeed captures correctly the leading contribution to the black hole shadow. We note that this is in particular also the case for the Dymnikova black hole, in which case approximations of the shadow based on finite orders of the Padé approximant are in general not expected to be analytic in $\mathfrak{c}$ (or $\epsilon$). While therefore intermediate results (notably table~\ref{Tab:ValuePade}) cannot be directly used, the final result in (\ref{CompactCorrectionShadow}) seems to correctly re-package the leading contributions in an effective manner.

%%%%%%%%%%%%%%%%%%%%%%%%%%%%%%%%%%%

%Particular attention must be paid to Dymnikova spacetime because the deformation function is not analytic around $\eta=0$.  We have to resort to the double expansion in power series of $u=e^{-1/\eta}$ and $\log u$. In fact, the outer event horizon can be expanded as $ \xh^\text{(Dymn)}=1-u+3u^2 \log u + \order{u^3}=1-\epsilon+\epsilon^2 +\order{\epsilon^3}$, while the photon sphere radius is
%\begin{equation}\label{dymn exact xps expansion}
%    \xps^\text{(Dymn)}=\frac{3}{2}\left(1 + \left( \sigma_1 + \lambda_1 \log \epsilon\right)\epsilon^{27/8} + \mathcal{O}( \epsilon^{33/8})\right) \ ,
%\end{equation}
%however, using the Padé approximant of the metric, we have to expand retaining lower powers in $\epsilon$, namely
%\begin{equation}\label{dymn xps expansion}
%    \xps^\text{(Dymn, Padé)}=\frac{3}{2}\left(1 + \left( \sigma_1 + \lambda_1 \log \epsilon\right)\epsilon + \mathcal{O}( \epsilon \log^2 \epsilon)\right) \ ,
%\end{equation}
%where $\sigma_1$ and $\lambda_1$ are analytically computed and given in Table \ref{subs example models}. On the other hand, the values that are found with the Padé approximation are $\sigma_1=-1$ for all orders and for $\lambda_1$ they are shown in Table on the right in Table \ref{HaywardDymntable}.
%\usepackage{booktabs,subcaption} % Assicurati di caricare questi pacchetti

\subsection{Dilaton black hole}\label{Sect:DilatonBH}
The black hole models of the previous Subsection have each been characterised by a single deformation function (\emph{i.e.} these examples satisfy $f=h$ in the metric (\ref{MetricGeneral})). Here we present a different example, which is not bound by this limitation. Indeed, in \cite{Garcia95,dilaton} a \emph{dilaton-axion} system has been studied with a metric resembling a black hole. In absence of the axion field (and without spin), the line element of the latter is given
\begin{equation}
    \dd s^2 = - \frac{\rd - 2 \mu}{\rd + 2 b} \dd t^2 + \frac{\rd + 2 b}{\rd - 2 \mu} \dd \rd^2 + (\rd^2 + 2 b \rd) \dd \Omega_2 \qq{with} r^2 = \rd^2 + 2 b \rd \qq{and} M=\mu + b \ .\label{AxioDil}
\end{equation}
After changing to the same coordinates as in (\ref{MetricGeneral}) we find (with $r=2\ma x$)
\begin{align}
&h=1+\frac{\eta-\sqrt{\eta^2+4x^2(1+\eta)^2}}{2x^2(1+\eta)}\,,&&f=\frac{\eta^2+4x^2(1+\eta)^2}{4x^2(1+\eta)^2}\,\left[1+\frac{\eta-\sqrt{\eta^2+4x^2(1+\eta)^2}}{2x^2(1+\eta)}\right]\,,&&\eta=\frac{b}{\mu}=\epsilon(2+\epsilon)\,,\label{MetDilatonBH}
\end{align}
where $\epsilon$ is defined as in (\ref{EpsEq}). For the coefficients (\ref{RescaledMetricFunctions}) we find
\begin{align}
&\mathfrak{t}_2=\frac{\epsilon  (\epsilon +2) (\epsilon  (\epsilon +2)+2) \left(\sqrt{(\epsilon  (\epsilon +2)+2)^2}-\epsilon  (\epsilon +2)\right)}{16 (\epsilon +1)   \sqrt{(\epsilon  (\epsilon +2)+2)^2}}=\frac{\epsilon}{4}-\frac{\epsilon^2}{8}+\mathcal{O}(\epsilon^3)\,,\hspace{1cm}\mathfrak{t}_4=-\frac{\epsilon}{12}+\frac{\epsilon^2}{96}+\mathcal{O}(\epsilon^3)\,,\nonumber\\   
&\mathfrak{x}_2=\frac{\epsilon  (\epsilon +2) (\epsilon  (\epsilon +2)+2) \left(3 \epsilon  (\epsilon +2) \left(\epsilon  (\epsilon +2)-\sqrt{(\epsilon  (\epsilon
   +2)+2)^2}+2\right)+4\right)}{64 (\epsilon +1)^3} =\frac{\epsilon}{4}-\frac{3\epsilon^2}{8}+\mathcal{O}(\epsilon^3) \,,\nonumber\\
&\mathfrak{x}_4=-\frac{\epsilon}{12}+\frac{5\epsilon^2}{32}+\mathcal{O}(\epsilon^3)\,,
\end{align}
\begin{figure}[t]
    \centering
    \includegraphics[width=0.8\textwidth]{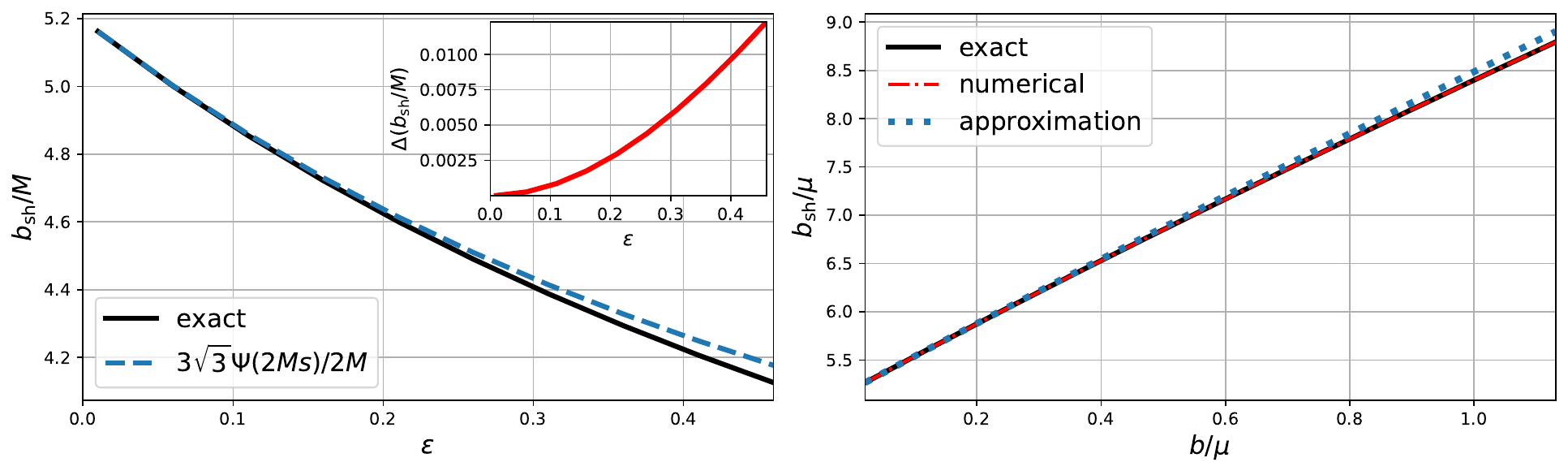}
    \caption{Left panel: Comparison of the shadow computed from eq.~(\ref{CompactCorrectionShadow}) (dashed blue line) to the exact numerical value (solid black line) for the dilaton black hole (\ref{MetDilatonBH})) as a function of $\epsilon$. The inset plot shows the relative difference between these two quantities. Right panel: Comparison to the analytic expression (\ref{DilatonAnalyticShadow}) of the shadow (solid black line) with a numerical computation (red line) and the approximation based on eq.~(\ref{CompactCorrectionShadow}) (dashed blue line), for a large range of the parameter $b/\mu$ defined in (\ref{MetDilatonBH}).}
    \label{fig: dilaton plots}
\end{figure}
which indeed are all of order $\mathcal{O}(\epsilon)$ and are thus compatible with the scaling (\ref{IllumScaling}). We can thus directly use Tables~\ref{Tab:PositionPade} and \ref{Tab:ValuePade} to calculate the position of the photon radius and the shadow, as shown in the following table
\begin{center}
\begin{tabular}{|c|c||c|c|c|}\hline
&&&\\[-10pt]
$N$ & $M$ & $\sigma_1\,\mathfrak{c}$ & $\tau_1\,\mathfrak{c}$ \\[2pt]\hline
&&&\\[-8pt]
2 & 1 & $-\frac{2}{3}\,(2+\sqrt{3})\,\epsilon+\mathcal{O}(\epsilon^2)$ &  $-\frac{4}{3}\,(1+\sqrt{3})\,\epsilon+\mathcal{O}(\epsilon^2)$\\[4pt]\hline
%%%
&&&\\[-10pt]
2 & 2 & $-\frac{6}{13}\,(4+\sqrt{3})\,\epsilon+\mathcal{O}(\epsilon^2)$ & $-\frac{4}{3}\,\sqrt{\frac{10}{3}+2\sqrt{3}}\,\epsilon +\mathcal{O}(\epsilon^2)$\\[4pt]\hline
   %%%
&&&\\[-8pt]
2 & 3 & $-\frac{8}{3}\,\epsilon+\mathcal{O}(\epsilon^2)$ & $-2\,\sqrt{3}\,\epsilon+\mathcal{O}(\epsilon^2)$ \\[4pt]\hline
   %%%
&&&\\[-8pt]
3 & 3 & $-\frac{8}{3}\,\epsilon+\mathcal{O}(\epsilon^2)$ & $-2\,\sqrt{3}\,\epsilon+\mathcal{O}(\epsilon^2)$\\[4pt]\hline
\end{tabular}
\end{center}
We note that this table is in fact identical with the results in the case of the Simpson-Visser II black hole. Indeed, this is due to the fact that
\begin{align}
&\mathfrak{t}_{2n}^{\text{axion-dilaton}}-\mathfrak{t}_{2n}^{\text{Simpson-Visser II}}=\mathcal{O}(\epsilon^3)\,,&&\text{and} &&\mathfrak{x}_{2n}^{\text{axion-dilaton}}-\mathfrak{t}_{2n}^{\text{Simpson-Visser II}}=\mathcal{O}(\epsilon^2)\,,\label{DilatonSimilarity}
\end{align}
thus leading to the identical results to leading order in $\epsilon$. We also remark that an analytic expression for the black hole shadow for the geometry (\ref{AxioDil}) was found in \cite{Wei:2013kza}
\begin{align}
\frac{\bsh}{\ma}=\frac{1}{\sqrt{2}(1+\eta)}\,\sqrt{27 + 36 \eta + 8 \eta^2 + (9 + 8 \eta)^{3/2}}=3\sqrt{3}-2\sqrt{3}\,\epsilon+\mathcal{O}(\epsilon)\,,\label{DilatonAnalyticShadow}
\end{align}
which indeed agrees with the result obtained from the approximations of Table~\ref{Tab:ValuePade} for sufficiently large $(N,M)$. The shadow has also been computed numerically in \cite{Rezzolla:2014mua} using different effective approximations. We also remark that (\ref{CompactCorrectionShadow}) gives an excellent approximation for the shadow for a large range of $\epsilon$, as is showcased in Figure~\ref{fig: dilaton plots}.

\section{Conclusions}\label{Sect:Conclusions}
In this paper we demonstrate how to use Padé approximants to extend the range of applicability of effective metric descriptions (EMDs) of quantum deformed black holes, beyond the immediate vicinity of the horizon. Following the approach advocated in \cite{Binetti:2022xdi,DelPiano:2023fiw,DelPiano:2024gvw}, we consider spherically symmetric and static black hole geometries, whose deformation with respect to the classical Schwarzschild metric is described by two functions (see (\ref{hfgen})) of the physical proper distance $\rho$ to the horizon. In order to solve for this distance in a self-consistent manner, in \cite{DelPiano:2023fiw} these two functions have been written in terms of Taylor series expansions (see  eq.~(\ref{SeriesExpansionMetric})) and therefore a concrete model of a black hole is defined through the effective coefficients $\{\theta_{2n}\}_{n\in\mathbb{N}^*}$ and $\{\xi_{2n}\}_{n\in\mathbb{N}^*}$. However, at regions outside of the immediate vicinity of the horizon, the Taylor series (\ref{SeriesExpansionMetric}) may converge very slowly (and thus require to know many of the effective coefficients) or not at all (depending on its radius of convergence). This therefore poses problems when trying to use the effective approach of \cite{DelPiano:2023fiw} to compute observables at a certain distance to the black hole, such as the black hole shadow (see Section~\ref{Sect:PhotonRadius} for the definition).\\

In this paper we therefore approximate the series expansions~(\ref{SeriesExpansionMetric}) by Padé approximants (see Appendix~\ref{App:DefPadeApproximant}) for the definition), which are not bound by a finite radius of convergence and are expected to provide a viable approximation to the black hole metric in a larger region of space-time. Starting with approximants of the metric deformation functions (see Section~\ref{Sect:PadeMetric}), the order $(N,M)$ determines how many of the initially infinitely many physical parameters $\{\theta_{2n}\}_{n\in\mathbb{N}^*}$ and $\{\xi_{2n}\}_{n\in\mathbb{N}^*}$ remain as effective parameters, while all remaining ones are replaced by functions of them (see (\ref{ReplacementCoefficients})). This concept of effective order of the approximation also percolates to the computation of physical observables, such as the physical distance (which is required for a self-consistent description of the metric) and notably the radius of the photon ring and the black hole shadow. In fact, in Section~\ref{Sect:PadePhoton}, we explain how to compute Padé approximants of the potential in order to determine the photon radius in a way that is compatible with a chosen order. Assuming furthermore that the quantum corrections are small and that deformations appear as analytic functions of a suitable small parameter, this allows us to provide effective expressions for the black hole shadow as a function of finitely many effective parameters. Indeed, taking the correction of the horizon position as an effective expansion parameter, we provide in Table~\ref{Tab:ValuePade} the leading quantum corrections to the black hole shadow as a function of the (rescaled) physical deformation parameters of the black hole. \\

By analysing the form of this effective result to order $(N,3)$ (with $N\in\mathbb{N}$), we find a particular pattern in which the physical deformation parameters contribute to the leading deformation of the shadow. This pattern allows us to conjecture its asymptotic form (\emph{i.e.} for $N\to \infty$), which we can furthermore resum in terms of the full metric deformation function in eq.~(\ref{CompactCorrectionShadow}). This leads us to the following conjecture for the effective black hole shadow, quantum corrected to leading order
\begin{align}
&\bsh\sim \frac{3\sqrt{3}}{2}\,\Psi(2\ma\mathfrak{s})\,,&&\text{with} &&\mathfrak{s}= \frac{1}{2}\,(\sqrt{3}+2\text{arccoth}\sqrt{3})\sim 1.5245\,,\label{EffectivePhotonSphere}
\end{align}
where $2\ma \mathfrak{s}$ is the physical distance of the classical photon sphere. The result (\ref{EffectivePhotonSphere}) takes into account the fully deformed metric (in the form of the deformation function $\Psi$), however, is an approximation in that it only captures the leading quantum correction to the shadow. While in intermediate steps we have parametrised the size of the quantum corrections through the deformation of the event horizon, the result (\ref{EffectivePhotonSphere}) does not depend on the details of this definition: we have indeed tested its robustness under alternative choices for gauging the size of quantum effects. Furthermore, preliminary studies suggest that corrections beyond (\ref{EffectivePhotonSphere}) are no longer linear in the functions $\Phi$ and $\Psi$ (or their derivatives). However, studying our effective result in numerous simple models of deformed black hole metrics also suggests that (\ref{EffectivePhotonSphere}) already provides a very good approximation.\\

Indeed, in Section~\ref{subs example models} we have illustrated our approach at the hand of a number of quantum deformed black holes that have been previously proposed in the literature~\cite{Hayward_2006,Simpson:2018tsi,Simpson:2019mud,Dymnikova:1992ux,Garcia95}. These examples can be cast in a form such that quantum corrections are indeed captured by a small parameter $\epsilon$, which is related to the deviation of the horizon position relative to the (classical) Schwarzschild case. The examples have been selected such that the metric deformations have functionally different dependencies on this parameter, ranging from rational functions~\cite{Hayward_2006} to a non-analytic behaviour (\emph{i.e.} in the form of $\log\epsilon$ contributions~\cite{Dymnikova:1992ux}). We have showcased our general reasoning by comparing effective descriptions in all these cases at different orders of the Padé approximant to the (numerically) exact values. In all cases, the Padé approximated metrics and observables compare very favourably to the expected results. Most importantly, even in cases where the deformation functions are not fully analytic in $\epsilon$, the effective shadow (\ref{EffectivePhotonSphere}) provides an approximation with an accuracy of $<1\%$ (for small enough values of $\epsilon$). This is much less than the experimental uncertainty (which is of the order of $17\%$). We can therefore turn measurements such as \cite{EventHorizonTelescope:2021dqv} for M$87^*$ around to provide a model independent upper bound on the size of the metric deformation function (see (\ref{ExperimentalBound})). \\

Our results show that the EMDs developed theoretically in previous works are powerful tools to study black hole observables that will be systematically explored by astrophysical experiments in the following years, with ever growing accuracy. We therefore expect our approach to provide a universal, model-independent framework to analyse such experimental results and compare them among each other. To further hone this analytical tool in the future, we plan to extend our results to larger classes of black holes, which notably carry (electric) charges and angular momentum. Similarly, we plan to consider new types of observables, that may capture the movement of (charged) matter in the vicinity of the black hole.

\subsection*{Acknowledgments}
We thank Vittorio De Falco, Maria Felicia De Laurentis,  Mikolaj Myszkowski, Mattia Paciarini, Luciano Rezzolla and Vania Vellucci for helpful discussions. S.H. would like to thank the Quantum Theory Center (QTC) at the Danish Institute for Advanced Study and IMADA of the University of Southern Denmark for the hospitality during the completion of this work. The work of F.S. is partially supported by the Carlsberg Foundation, grant CF22-0922.

\appendix

%%%%%%%%%%%%%%%%%%%%%%%%%

%%%%%%%%%%%%%%%%%%%%%%%%%

%%%%%%%%%%%%%%%%%%%%%%%%%
\section{Convergence}\label{Sect:Convergence}
\begin{figure}[!h]
    \centering
    \includegraphics[width=\textwidth]{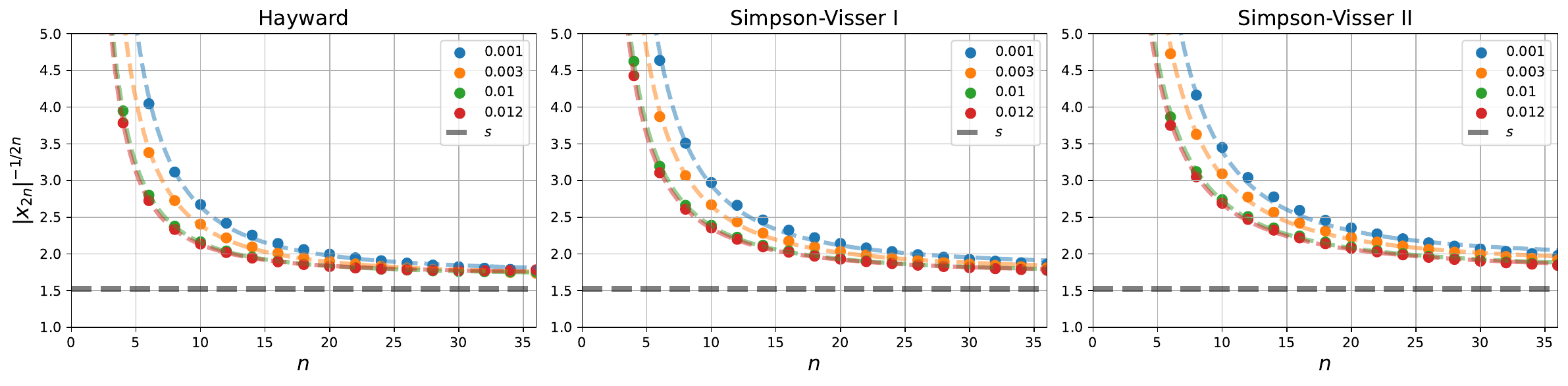}
    \caption{Asymptotic behaviour of $|\mathfrak{x}_{2n}|^{-\frac{1}{2n}}$, respectively from the left, in the case of the Hayward, Simpson-Visser I and Simpson-Visser II black holes. The dashed coloured lines are interpolations of the form $a+\frac{b}{x^c}$, while the dashed black line is the value of $\mathfrak{s}$ in (\ref{Svalue}). }
    \label{fig:ConvergenceExamples}
\end{figure}

\begin{figure}[!h]
\centering
\includegraphics[width=0.4\textwidth]{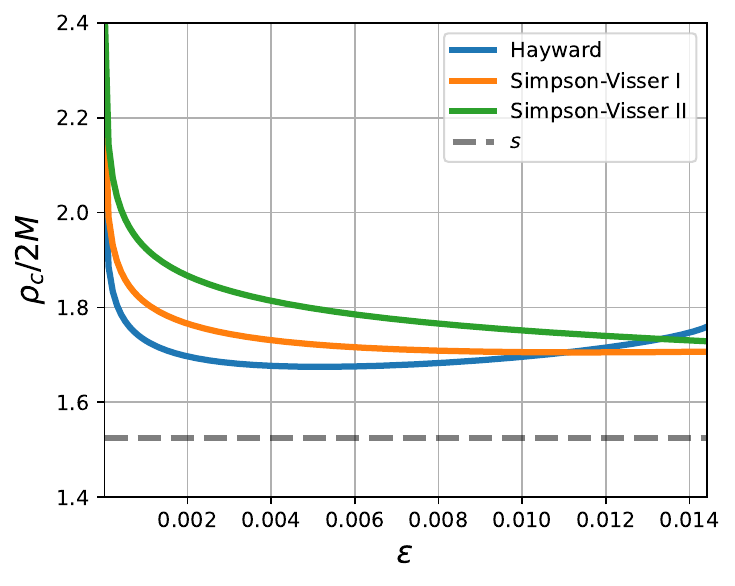}
\caption{Numerical estimate of the radius of convergence in the case of the Hayward, Simpson-Visser I and Simpson-Visser II black hole. The dashed black line is the value of $\mathfrak{s}$ in (\ref{Svalue}).}
   \label{fig:ShadowLimit}
\end{figure}

In order to obtain the effective expression (\ref{CompactCorrectionShadow}) of the black hole shadow in Section~\ref{Sect:PadePhoton}, we have implicitly assumed that the Taylor series (\ref{SeriesExpansionMetric}) of the metric deformation function $\Psi$ converges at $\rho=2\ma\mathfrak{s}$. To get an intuition whether this is indeed the case, we study numerically the radii of convergence for the Hayward, Simpson-Visser I and II black hole. Indeed, the coefficients $|\mathfrak{x}_{2n}|^{-\frac{1}{2n}}$ up to $n=36$ are plotted for some values of $\epsilon$ in the three panels in Figure~\ref{fig:ConvergenceExamples}. The dashed lines in each case represent numerical interpolations of the form $a+\frac{b}{x^c}$ (with $a,b,c\in\mathbb{R}$ and $c>0$). The value of $a$ in each case is a numerical approximation of the radius of convergence $\rho_c$ of the function $\Psi$. Based on a more systematic study, this radius is plotted on the right in Figure~\ref{fig:ShadowLimit} as a function of $\epsilon$, with the dashed line representing the value of $\mathfrak{s}$. As is evident, at least for small values of $\epsilon$, $\mathfrak{s}$ lies within the radius of convergence of the series, thus justifying the resummation leading to~(\ref{CompactCorrectionShadow}). Finally, due to the inherent similarity between the axion-dilaton black hole and the Simpson-Visser II geometry (see (\ref{DilatonSimilarity})) at leading order in $\epsilon$, we expect a similar result also hold for the example discussed in Section~\ref{Sect:DilatonBH}.

%\begin{figure}[!h]
%    \centering
%    \includegraphics[width=0.3\textwidth]{images/ConvergencePointsHayward.png}\hspace{0.25cm} \includegraphics[width=0.3\textwidth]{images/ConvergencePointsSV1.png}\hspace{0.25cm} \includegraphics[width=0.3\textwidth]{images/ConvergencePointsSV2.png}
%    \caption{Asymptotic behaviour of $\mathfrak{x}_{2n}|^{\frac{1}{2n}}$ in the case of the Hayward (left), Simpson-Visser I (middle) and Simpson-Visser II (right) black hole. The dashed coloured lines are interpolations of the form $a+\frac{b}{x^c}$, while the dashed black line is the value of $\mathfrak{s}$ in (\ref{Svalue}).}
%    \label{fig:ConvergenceExamples}
%\end{figure}

%%%%%%%%%%%%%%%%%%%%%%%%%%%%

\section{Padè Approximant of a Series Expansion}\label{App:DefPadeApproximant}
Let $U:\,[0,x_0]\rightarrow \mathbb{R}$ be a function with series expansion
\begin{align}
U(x)=\sum_{k=0}^\infty  \mathfrak{u}_k\,x^k\,,
\end{align}
that has radius of convergence $x_0>0$. We then introduce the \emph{Padé approximant} \cite{bender78:AMM} as the quotient of two polynomials 
\begin{align} \label{padedef}
&U_{N,M}:=\frac{P_N(x)}{Q_M(x)}\,,&&\text{with} &&\begin{array}{l}P_N=\sum_{n=0}^N p_n\,x^n\,,\\ Q_M=1+\sum_{m=1}^Mq_m\,x^m\,,\end{array}
\end{align}
such that
\begin{align}
U_{N,M}(x)=\sum_{k=0}^{N+M}\mathfrak{u}_k\,x^k+\mathcal{O}(x^{N+M+1})\,.
\end{align}
If $U_{N,M}$ exists, the coefficients $\{p_n\}$ and $\{q_m\}$ are unique and given by the linear equations
\begin{align}
&\mathcal{M}_1\cdot\left(\begin{array}{c}q_M \\ q_{M-1}\\ \vdots \\ q_1\end{array}\right)=-\left(\begin{array}{c}\mathfrak{u}_{N+1} \\ \mathfrak{u}_{N+2}\\ \vdots \\ \mathfrak{u}_{N+M}\end{array}\right)\,,&&\text{with} &&\mathcal{M}_1=\left(\begin{array}{cccc} \mathfrak{u}_{N-M+1} & \mathfrak{u}_{N-M+2} & \cdots   & \mathfrak{u}_{N} \\ \mathfrak{u}_{N-M+2} & \mathfrak{u}_{N-M+3} & \cdots & \mathfrak{u}_{N+1} \\ \vdots & \vdots  & & \vdots \\ \mathfrak{u}_{N} & \mathfrak{u}_{N+1} &\cdots  & \mathfrak{u}_{N+M-1} \end{array}\right)\,,\nonumber\\
&\mathcal{M}_2\cdot\left(\begin{array}{c}1 \\ q_{1}\\ \vdots \\ q_M\end{array}\right)=\left(\begin{array}{c} p_0 \\ p_1 \\  \vdots \\ p_N\end{array}\right)\,,&&\text{with} &&\mathcal{M}_2=\left(\begin{array}{ccccccc} 0 & 0 & \cdots   & & & & 0 \\ \mathfrak{u}_1 & 0 & 0 & \cdots   & & & 0 \\ \mathfrak{u}_2 & \mathfrak{u}_1 & 0 & 0 &\cdots   & & 0 \\ \vdots & \vdots & \vdots & \vdots & \ddots & &\vdots \\ \mathfrak{u}_N & \mathfrak{u}_{N-1} & \mathfrak{u}_{N-2} & \cdots &  & \mathfrak{u}_{N-M-1} &\mathfrak{u}_{N-M} \end{array}\right)\,,
\end{align}
If it exists, an explicit form of $U_{N,M}$ can be given in the form
\begin{align}
U_{N,M}(x)=\frac{\left|\begin{array}{cccc}\mathfrak{u}_{N-M+1} & \mathfrak{u}_{N-M+2} & \cdots & \mathfrak{u}_{N+1} \\ \vdots & \vdots & \ddots & \vdots \\ \mathfrak{u}_N & \mathfrak{u}_{N+1} & \cdots & \mathfrak{u}_{N+M} \\ \sum_{j=M}^N \mathfrak{u}_{j-M}x^j & \sum_{j=M-1}^N \mathfrak{u}_{j-M+1}x^j & \cdots & \sum_{j=0}^N \mathfrak{u}_j x^j\end{array}\right|}{\left|\begin{array}{cccc}\mathfrak{u}_{N-M+1} & \mathfrak{u}_{N-M+2} & \cdots & \mathfrak{u}_{N+1} \\ \vdots & \vdots & \ddots & \vdots \\ \mathfrak{u}_{N} & \mathfrak{u}_{N+1} & \cdots & \mathfrak{u}_{N+M} \\ x^M & x^{M-1} & \cdots & 1\end{array}\right|}\,.
\end{align}

\bibliographystyle{unsrt}
\bibliography{bibliography.bib}

\begin{thebibliography}{10}

\bibitem{DelPiano:2023fiw}
Manuel Del~Piano, Stefan Hohenegger, and Francesco Sannino.
\newblock {Quantum black hole physics from the event horizon}.
\newblock {\em Phys. Rev. D}, 109(2):024045, 2024.

\bibitem{DelPiano:2024gvw}
Manuel Del~Piano, Stefan Hohenegger, and Francesco Sannino.
\newblock {Effective metric descriptions of quantum black holes}.
\newblock {\em Eur. Phys. J. C}, 84(12):1273, 2024.

\bibitem{Schwarzschild:1916uq}
Karl Schwarzschild.
\newblock {On the gravitational field of a mass point according to Einstein's theory}.
\newblock {\em Sitzungsber. Preuss. Akad. Wiss. Berlin (Math. Phys. )}, 1916:189--196, 1916.

\bibitem{Wald:1984rg}
Robert~M. Wald.
\newblock {\em {General Relativity}}.
\newblock Chicago Univ. Pr., Chicago, USA, 1984.

\bibitem{Maggiore:2018sht}
Michele Maggiore.
\newblock {\em {Gravitational Waves. Vol. 2: Astrophysics and Cosmology}}.
\newblock Oxford University Press, 3 2018.

\bibitem{Chandrasekhar:1985kt}
Subrahmanyan Chandrasekhar.
\newblock {\em {The mathematical theory of black holes}}.
\newblock 1985.

\bibitem{Carroll:2004st}
Sean~M. Carroll.
\newblock {\em {Spacetime and Geometry}}.
\newblock Cambridge University Press, 7 2019.

\bibitem{misner1973gravitation}
Charles~W Misner, Kip~S Thorne, and John~Archibald Wheeler.
\newblock {\em Gravitation}.
\newblock Macmillan, 1973.

\bibitem{EventHorizonTelescope:2019dse}
Kazunori Akiyama et~al.
\newblock {First M87 Event Horizon Telescope Results. I. The Shadow of the Supermassive Black Hole}.
\newblock {\em Astrophys. J. Lett.}, 875:L1, 2019.

\bibitem{EventHorizonTelescope:2021dqv}
Prashant Kocherlakota et~al.
\newblock {Constraints on black-hole charges with the 2017 EHT observations of M87*}.
\newblock {\em Phys. Rev. D}, 103(10):104047, 2021.

\bibitem{EventHorizonTelescope:2022wkp}
Kazunori Akiyama et~al.
\newblock {First Sagittarius A* Event Horizon Telescope Results. I. The Shadow of the Supermassive Black Hole in the Center of the Milky Way}.
\newblock {\em Astrophys. J. Lett.}, 930(2):L12, 2022.

\bibitem{EventHorizonTelescope:2022xqj}
Kazunori Akiyama et~al.
\newblock {First Sagittarius A* Event Horizon Telescope Results. VI. Testing the Black Hole Metric}.
\newblock {\em Astrophys. J. Lett.}, 930(2):L17, 2022.

\bibitem{Vagnozzi:2022moj}
Sunny Vagnozzi et~al.
\newblock {Horizon-scale tests of gravity theories and fundamental physics from the Event Horizon Telescope image of Sagittarius A}.
\newblock {\em Class. Quant. Grav.}, 40(16):165007, 2023.

\bibitem{LIGOScientific:2020tif}
R.~Abbott et~al.
\newblock {Tests of general relativity with binary black holes from the second LIGO-Virgo gravitational-wave transient catalog}.
\newblock {\em Phys. Rev. D}, 103(12):122002, 2021.

\bibitem{LIGOScientific:2020stg}
R.~Abbott et~al.
\newblock {GW190412: Observation of a Binary-Black-Hole Coalescence with Asymmetric Masses}.
\newblock {\em Phys. Rev. D}, 102(4):043015, 2020.

\bibitem{LIGOScientific:2018dkp}
B.~P. Abbott et~al.
\newblock {Tests of General Relativity with GW170817}.
\newblock {\em Phys. Rev. Lett.}, 123(1):011102, 2019.

\bibitem{LIGOScientific:2016dsl}
B.~P. Abbott et~al.
\newblock {Binary Black Hole Mergers in the first Advanced LIGO Observing Run}.
\newblock {\em Phys. Rev. X}, 6(4):041015, 2016.
\newblock [Erratum: Phys.Rev.X 8, 039903 (2018)].

\bibitem{LIGOScientific:2016aoc}
B.~P. Abbott et~al.
\newblock {Observation of Gravitational Waves from a Binary Black Hole Merger}.
\newblock {\em Phys. Rev. Lett.}, 116(6):061102, 2016.

\bibitem{Cadoni:2022vsn}
M.~Cadoni, M.~De~Laurentis, I.~De~Martino, R.~Della~Monica, M.~Oi, and A.~P. Sanna.
\newblock {Are nonsingular black holes with super-Planckian hair ruled out by S2 star data?}
\newblock {\em Phys. Rev. D}, 107(4):044038, 2023.

\bibitem{Binetti:2022xdi}
Emanuele Binetti, Manuel Del~Piano, Stefan Hohenegger, Franco Pezzella, and Francesco Sannino.
\newblock {Effective theory of quantum black holes}.
\newblock {\em Phys. Rev. D}, 106(4):046006, 2022.

\bibitem{DAlise:2023hls}
Alessandra D'Alise, Giuseppe Fabiano, Domenico Frattulillo, Stefan Hohenegger, Davide Iacobacci, Franco Pezzella, and Francesco Sannino.
\newblock {Positivity conditions for generalized Schwarzschild space-times}.
\newblock {\em Phys. Rev. D}, 108(8):084042, 2023.

\bibitem{Li:2023djs}
Yong-Zhuang Li and Xiao-Mei Kuang.
\newblock {Precession of bounded orbits and shadow in quantum black hole spacetime}.
\newblock {\em Phys. Rev. D}, 107(6):064052, 2023.

\bibitem{Belfiglio:2024qsa}
Alessio Belfiglio, Orlando Luongo, Stefano Mancini, and Sebastiano Tomasi.
\newblock {Entanglement entropy in quantum black holes}.
\newblock 3 2024.

\bibitem{Perlick:2021aok}
Volker Perlick and Oleg~Yu. Tsupko.
\newblock {Calculating black hole shadows: Review of analytical studies}.
\newblock {\em Phys. Rept.}, 947:1--39, 2022.

\bibitem{Bozza:2002zj}
V.~Bozza.
\newblock {Gravitational lensing in the strong field limit}.
\newblock {\em Phys. Rev. D}, 66:103001, 2002.

\bibitem{Hayward_2006}
Sean~A. Hayward.
\newblock Formation and evaporation of nonsingular black holes.
\newblock {\em Physical Review Letters}, 96(3), 1 2006.

\bibitem{Simpson:2018tsi}
Alex Simpson and Matt Visser.
\newblock {Black-bounce to traversable wormhole}.
\newblock {\em JCAP}, 02:042, 2019.

\bibitem{Simpson:2019mud}
Alex Simpson and Matt Visser.
\newblock {Regular black holes with asymptotically Minkowski cores}.
\newblock {\em Universe}, 6(1):8, 2019.

\bibitem{Dymnikova:1992ux}
I.~Dymnikova.
\newblock {Vacuum nonsingular black hole}.
\newblock {\em Gen. Rel. Grav.}, 24:235--242, 1992.

\bibitem{Garcia95}
Alberto Garc\'{\i}a, Dmitri Galtsov, and Oleg Kechkin.
\newblock Class of stationary axisymmetric solutions of the einstein-maxwell-dilaton-axion field equations.
\newblock {\em Phys. Rev. Lett.}, 74:1276--1279, Feb 1995.

\bibitem{Synge1966}
J.~L. Synge.
\newblock {The Escape of Photons from Gravitationally Intense Stars}.
\newblock {\em Monthly Notices of the Royal Astronomical Society}, 131(3):463--466, 02 1966.

\bibitem{Perlick:2015vta}
Volker Perlick, Oleg~Yu. Tsupko, and Gennady~S. Bisnovatyi-Kogan.
\newblock {Influence of a plasma on the shadow of a spherically symmetric black hole}.
\newblock {\em Phys. Rev. D}, 92(10):104031, 2015.

\bibitem{Gan:2021pwu}
Qingyu Gan, Peng Wang, Houwen Wu, and Haitang Yang.
\newblock {Photon spheres and spherical accretion image of a hairy black hole}.
\newblock {\em Phys. Rev. D}, 104(2):024003, 2021.

\bibitem{Cunha:2020azh}
Pedro V.~P. Cunha and Carlos A.~R. Herdeiro.
\newblock {Stationary black holes and light rings}.
\newblock {\em Phys. Rev. Lett.}, 124(18):181101, 2020.

\bibitem{dilaton}
Alberto Garc\'{\i}a, Dmitri Galtsov, and Oleg Kechkin.
\newblock Class of stationary axisymmetric solutions of the einstein-maxwell-dilaton-axion field equations.
\newblock {\em Phys. Rev. Lett.}, 74:1276--1279, Feb 1995.

\bibitem{Rezzolla:2014mua}
Luciano Rezzolla and Alexander Zhidenko.
\newblock {New parametrization for spherically symmetric black holes in metric theories of gravity}.
\newblock {\em Phys. Rev. D}, 90(8):084009, 2014.

\bibitem{Culetu:2014lca}
Hristu Culetu.
\newblock {On a regular charged black hole with a nonlinear electric source}.
\newblock {\em Int. J. Theor. Phys.}, 54(8):2855--2863, 2015.

\bibitem{bardeen1968}
James {Bardeen}.
\newblock {Non-singular general relativistic gravitational collapse}.
\newblock In {\em Proceedings of the 5th International Conference on Gravitation and the Theory of Relativity}, page~87, September 1968.

\bibitem{Frolov:PhysRevD.94.104056}
Valeri~P. Frolov.
\newblock Notes on nonsingular models of black holes.
\newblock {\em Phys. Rev. D}, 94:104056, Nov 2016.

\bibitem{Wei:2013kza}
Shao-Wen Wei and Yu-Xiao Liu.
\newblock {Observing the shadow of Einstein-Maxwell-Dilaton-Axion black hole}.
\newblock {\em JCAP}, 11:063, 2013.

\bibitem{bender78:AMM}
C.~M. Bender and S.~A. Orszag.
\newblock {\em {Advanced Mathematical Methods for Scientists and Engineers}}.
\newblock McGraw-Hill, 1978.

\end{thebibliography}
\end{document}